\documentclass[12pt]{article}
\pdfoutput=1

\usepackage{amsmath}
\usepackage{amssymb}
\usepackage{graphicx}
\usepackage{cite}
\usepackage{multirow}
\usepackage{longtable}
\usepackage{lscape}
\usepackage{subfigure}

\newcommand{\be}{\begin{equation}}
\newcommand{\ee}{\end{equation}}
\newcommand{\ba}{\begin{eqnarray}}
\newcommand{\ea}{\end{eqnarray}}

\newcommand{\ms}{\Delta m^2_{21}}
\newcommand{\ma}{\Delta m^2_{31}}
\newcommand{\meff}{\Delta m^2_{\textrm{eff}}}
\newcommand{\mefft}{(\Delta m^2_{\textrm{eff}})^{\textrm{true}}}

\newcommand{\sch}{\sin^2\theta_{13}}
\newcommand{\stch}{\sin^22\theta_{13}}
\newcommand{\sa}{\sin^2\theta_{23}}
\newcommand{\sta}{\sin^22\theta_{23}}
\newcommand{\mst}{(\Delta m^2_{21})^{\textrm{true}}}

\newcommand{\ssst}{\sin^2\theta_{12}^{\textrm{true}}}
\newcommand{\scht}{\sin^2\theta_{13}^{\textrm{true}}}
\newcommand{\stcht}{\sin^22\theta_{13}^{\textrm{true}}}
\newcommand{\sat}{\sin^2\theta_{23}^{\textrm{true}}}
\newcommand{\stat}{\sin^22\theta_{23}^{\textrm{true}}}
\newcommand{\dcpt}{\delta_{\textrm{CP}}^{\textrm{true}}}
\newcommand{\dcp}{\delta_{\textrm{CP}}}
\newcommand{\tmt}{\theta_{23}}
\newcommand{\tet}{\theta_{13}}
\newcommand{\tem}{\theta_{12}}

\newcommand{\temt}{\theta_{12}^{\textrm{true}}}

\def\numu{{\nu_{\mu}}}

\newcommand{\capdef}{}
\newcommand{\mycaption}[2][\capdef]{\renewcommand{\capdef}{#2}
       \caption[#1]{{\footnotesize #2}}}
\makeatletter
\renewcommand{\fnum@table}{\textbf{\tablename~\thetable}}
\renewcommand{\fnum@figure}{\textbf{\figurename~\thefigure}}
\makeatother

\begin{document}

\begin{titlepage}

\renewcommand{\thefootnote}{\alph{footnote}}

\vspace*{-3.cm}
\begin{flushright}
IFIC/12-81 \\
CFTP/12-019 \\
\end{flushright}

\vspace*{0.2cm}

\renewcommand{\thefootnote}{\fnsymbol{footnote}}
\setcounter{footnote}{-1}

{\begin{center}

{\Large\bf
Exploring the Earth matter effect with \\ atmospheric neutrinos in ice
\\[0.3cm] 

}

\end{center}}

\renewcommand{\thefootnote}{\alph{footnote}}

\vspace*{.6cm}

{\begin{center} {
               \large{
                 Sanjib Kumar
                 Agarwalla\footnote[1]{\makebox[1.cm]{Email:} 
                 Sanjib.Agarwalla@ific.uv.es}},
               \large{
                 Tracey Li\footnote[2]{\makebox[1.cm]{Email:}
                 tracey.li@ific.uv.es}},
               \large{
                 Olga Mena\footnote[3]{\makebox[1.cm]{Email:}
                 omena@ific.uv.es}}
                 and
                 \large{
                 Sergio Palomares-Ruiz\footnote[4]{\makebox[1.cm]{Email:} 
                 sergio.palomares.ruiz@ist.utl.pt}}
                 }
\end{center}
}
\vspace*{0cm}
{\it
\begin{center}

\footnotemark[1]${}^,$\footnotemark[2]${}^,$\footnotemark[3]
       Instituto de F\'{\i}sica Corpuscular, CSIC-Universitat de
       Val\`encia, \\ 
       Apartado de Correos 22085, E-46071 Valencia, Spain \\
      
\vspace*{.2cm}       
       
\footnotemark[4]
       Centro de F\'{\i}sica Te\'{o}rica de Part\'{\i}culas (CFTP),
       Instituto Superior T\'{e}cnico, \\ 
       Universidade T\'{e}cnica de Lisboa, Av. Rovisco Pais 1,
       1049-001 Lisboa, Portugal  
       
\end{center}}
             
\vspace*{0.5cm}

{\Large \bf
\begin{center} Abstract \end{center}  }

We study the possibility to perform neutrino oscillation tomography and to determine the neutrino mass hierarchy in kilometer-scale ice \v{C}erenkov detectors by means of the $\tet$-driven matter effects which occur during the propagation of atmospheric neutrinos deep through the Earth.  We consider the ongoing IceCube/DeepCore neutrino observatory and future planned extensions, such as the PINGU detector, which has a lower energy threshold.  Our simulations include the impact of marginalization over the neutrino oscillation parameters and a fully correlated systematic uncertainty on the total number of events.  For the current best-fit value of the mixing angle $\tet$, the DeepCore detector, due to its relatively high-energy threshold, could only be sensitive to fluctuations on the normalization of the Earth's density of $\Delta \rho \simeq \pm 10\%$ at $\sim 1.6\sigma$~CL after 10~years in the case of a true normal hierarchy. For the two PINGU configurations we consider, overall density fluctuations of $\Delta \rho \simeq \pm 3\%$ ($\pm 2\%$) could be measured at the $2\sigma$~CL after 10~years, also in the case of a normal mass hierarchy.  We also compare the prospects to determine the neutrino mass hierarchy in these three configurations and find that this could be achieved at the $5\sigma$~CL, for both hierarchies, after 5~years in DeepCore and about 1~year in PINGU.  This clearly shows the importance of lowering the energy threshold below 10~GeV so that detectors are fully sensitive to the resonant matter effects.

\vspace*{.5cm}

\end{titlepage}
\newpage

\renewcommand{\thefootnote}{\arabic{footnote}}
\setcounter{footnote}{0}

\section{Introduction}
\label{sec:introduction}

Cosmic ray interactions in the atmosphere are a natural source of
neutrinos with baselines that span three orders of magnitude and
energies in the range from MeV to well above TeV.  This implies that
different scales and effects could be accessible by studying
atmospheric neutrinos.  In fact, the most important result of the
latest years in neutrino physics was the discovery in 1998 of
neutrino oscillations at the Super-Kamiokande detector by using
atmospheric neutrino data in the GeV energy range~\cite{Fukuda:1998mi}.

Among the potential effects that could be explored are the
resonant effects in the propagation of GeV atmospheric neutrinos
through the Earth that mainly affect the subleading $\nu_\mu
\rightarrow \nu_e$ and $\nu_e \rightarrow \nu_\mu$ (or $\bar\nu_\mu
\rightarrow \bar\nu_e$ and $\bar\nu_e \rightarrow \bar\nu_\mu$)
transitions.  When atmospheric neutrinos deeply cross the Earth's
mantle, the Mikheyev-Smirnov-Wolfenstein (MSW) resonance~\cite{Wolfenstein:1977ue, Mikheev:1986gs}
could be in action~\cite{Banuls:2001zn, Bernabeu:2001jb,
  Bernabeu:2002tj}.  On the other hand, when they also traverse the
core, resonant effects, different from the MSW resonance, could show
up~\cite{Petcov:1998su, Akhmedov:1998ui, Petcov:1998, Akhmedov:1998xq,
  Chizhov:1998ug, Chizhov:1999az, Chizhov:1999he, Akhmedov:2005yj,
  Akhmedov:2006hb}.  These effects are well known and their physics
potential in iron-magnetized calorimeter, water-\v{C}erenkov and
liquid argon detectors has been thoroughly studied in the
literature~\cite{Bernabeu:2001xn, GonzalezGarcia:2002mu, Bernabeu:2003yp, PalomaresRuiz:2003kz, PalomaresRuiz:2004tk, Petcov:2004bd, Indumathi:2004kd, GonzalezGarcia:2004cu, Gandhi:2004bj, Huber:2005ep, Choubey:2005zy, Petcov:2005rv, Indumathi:2006gr, Samanta:2006sj, Gandhi:2007td, Gandhi:2008zs, Samanta:2009qw, Samanta:2010xm, GonzalezGarcia:2011my, Blennow:2012gj, Barger:2012fx, Ghosh:2012}. 

It is very well known that the propagation of atmospheric neutrinos
with energies in the range of a few GeV deeply through the Earth is
sensitive to the currently unknown neutrino mass
hierarchy~\cite{Banuls:2001zn, Bernabeu:2001xn, Bernabeu:2003yp,
  Petcov:2004bd} (or neutrino mass ordering), i.e., whether the third
neutrino mass eigenstate is the lightest or the heaviest mass
eigenstate.  In the latter case, the hierarchy is called a
normal hierarchy (NH) and in the former case, inverted hierarchy (IH).  However, the ability to measure the neutrino mass hierarchy exploiting the resonant oscillation phenomena of atmospheric neutrinos inside the Earth crucially relies on the value of the mixing angle $\tet$.  Recent measurements of this angle from the Daya Bay~\cite{An:2012eh}, RENO~\cite{Ahn:2012nd} and Double Chooz~\cite{Abe:2012tg} reactor experiments, in addition to the 
long-baseline T2K experiment~\cite{Abe:2011sj}, seem to indicate that
$\tet \sim 9^\circ$~\cite{Tortola:2012te, Fogli:2012ua, GonzalezGarcia:2012sz}.  This large value of $\tet$ opens up the possibility of using the atmospheric neutrino fluxes not only to determine the neutrino mass hierarchy, but also to study features of
the Earth's matter density profile, by exploiting the $\tet$-driven matter effects. 

There are three different ways that have been proposed to infer some
information about the internal structure of the Earth by exploiting the
different effects of neutrino propagation in matter: neutrino
absorption tomography, neutrino oscillation tomography and neutrino
diffraction.  The idea of using the absorption of very high energy
neutrinos to explore the Earth's interior dates back to
1974~\cite{Volkova:1974} and is analogous to X--ray tomography, but instead exploits the attenuation of the neutrino flux for energies $E_\nu \gtrsim 10$~TeV~\cite{Gandhi:1995tf}.  There are numerous studies
which have considered neutrinos of different origins, such as man-made 
neutrinos~\cite{Volkova:1974, Nedyalkov:1981, Nedyalkov:1981yy,
  Nedyalkov:1983, DeRujula:1983ya, Wilson:1983an, Askarian:1985ca,
  Volkova:1985zc, Tsarev:1985, Borisov:1986sm, Tsarev:1986xg},
extraterrestrial neutrinos\footnote{This possibility was first
  suggested by J.~Learned and H.~Bradner in the late 1970s, but we
  lack a proper reference.}~\cite{Wilson:1983an, Kuo:1995,
  Crawford:1995, Jain:1999kp, Reynoso:2004dt} and atmospheric
neutrinos~\cite{GonzalezGarcia:2007gg, Borriello:2009ad,
  Takeuchi:2010, Romero:2011zzb}.  On the other hand, neutrino
oscillation tomography relies on the matter effects in neutrino
oscillations and it has been considered by studying man-made
beams~\cite{Ermilova:1986ph, Nicolaidis:1987fe, Ermilova:1988pw,
  Nicolaidis:1990jm, Ohlsson:2001ck, Ohlsson:2001fy, Winter:2005we,
  Minakata:2006am, Gandhi:2006gu, Tang:2011wn, Arguelles:2012nw},
solar~\cite{Ioannisian:2002yj, Akhmedov:2005yt} and supernova
neutrinos~\cite{Lindner:2002wm, Akhmedov:2005yt}.  The third
possibility is based on studying the diffraction pattern of
crystalline matter in the interior of the Earth caused by coherent
neutrino scattering, but it is technologically
unfeasible~\cite{Fortes:2006}.

In the case of atmospheric neutrinos, only neutrino absorption
tomography has been considered within the context of kilometer-scale
detectors~\cite{GonzalezGarcia:2007gg, Borriello:2009ad,
  Takeuchi:2010, Romero:2011zzb}.  However, resonant effects are very
sensitive to the matter density that neutrinos traverse, so one could
also think of doing neutrino oscillation tomography with atmospheric
neutrinos with energies in the GeV range\footnote{This was suggested
  in Ref.~\cite{Akhmedov:2006hb}, where the changes in the transition
  probabilities for different Earth density distributions were
  studied.}.  This is one of the main goals of this work and
represents a completely different technique to determine the Earth's
density distribution to those used in geophysics, mainly based on the
analysis of seismic waves.  Although, in principle, geophysics can
obtain more precise results, its inferences are based on numerous
assumptions because the velocities of seismic waves not only depend on the density, but also on the composition, temperature, pressure and elastic properties of the medium.  Thus, the results from atmospheric neutrino tomography would represent an independent and complementary assessment of the Earth's internal structure. 

Neutrino telescopes such as AMANDA~\cite{Ahrens:2002gq},
IceCube~\cite{Ahrens:2003ix} and Antares~\cite{Aslanides:1999vq} have
already collected a large number of atmospheric neutrino
events~\cite{Achterberg:2007bi, Abbasi:2009nfa, Abbasi:2010qv,
  Abbasi:2010ie, Abbasi:2011ui, Aguilar:2010kg,
  AdrianMartinez:2012ph}, even in spite of their high energy threshold
($\sim$~100~GeV) and the steeply falling atmospheric neutrino spectrum
($\sim E^{-3.7}_\nu$).  Whereas this is a very high threshold for studying neutrino oscillations with atmospheric neutrinos,  a new generation of neutrino telescopes with lower energy thresholds ($\sim$~10~GeV), such as the DeepCore extension of the Icecube detector, is currently taking data successfully~\cite{Collaboration:2011ym, IceCube:2011ah,
  Ha:2012np} and further natural extensions of this are being planned.
The Precision IceCube Next Generation Upgrade (PINGU) has been proposed
in order to reduce the detection threshold down to a few
GeV~\cite{Koskinen:2011zz}.  Although reaching those energy thresholds in these multi-Mton scale neutrino telescopes is very challenging, if successful, the atmospheric neutrino events detected at them would offer a great opportunity for detailed oscillation studies, including the determination of the neutrino mass hierarchy~\cite{Mena:2008rh, Akhmedov:2012ah}, tau neutrino appearance searches~\cite{Giordano:2010pr}, and precise measurements of the atmospheric neutrino oscillation parameters~\cite{FernandezMartinez:2010am}.  Given the large value of
$\tet$ measured by the different reactor~\cite{An:2012eh, Ahn:2012nd, Abe:2012tg} and the T2K~\cite{Abe:2011sj} experiments, the prospects look promising.  It therefore seems timely to assess their ability to infer some information about the Earth's matter density profile and how this could also affect the determination of the neutrino mass hierarchy from the atmospheric neutrino events.  In this work we study the future prospects for DeepCore and PINGU by considering the experimental capabilities of the two detectors and present the expected sensitivities to both of these observables.   
  
The structure of the paper is as follows.  We start in
Sec.~\ref{sec:theory} by revisiting the atmospheric neutrino fluxes, the Earth's internal structure and the matter effects which affect the
propagation of GeV neutrinos inside the Earth.  Sec.~\ref{sec:analysis} contains a description of the detectors
considered here, as well as of the $\chi^2$ analysis methods used to
calculate the sensitivities to the Earth's matter density and neutrino
mass hierarchy, which are presented in Sec.~\ref{sec:results}.  Finally, in Sec.~\ref{sec:conclusions} we draw our conclusions.

\section{Atmospheric neutrino fluxes and oscillations in the Earth}
\label{sec:theory}

\subsection{Atmospheric neutrino flux}

When cosmic rays traverse the Earth's atmosphere, hadronic showers are
produced by their interactions with the particles that constitute
it. Depending on the energy of the primary cosmic ray, different
secondaries can be produced.  Below $\sim$~100~GeV, the neutrino flux is dominated by the pion decay chain, whereas above these energies, kaon decays dominate neutrino production.  The decay chain is the following:
\begin{equation}
\begin{array}{lllll}
p+X&\rightarrow &\pi^{\pm}/K^{\pm}&+ & Y  \\
&   &\pi^{\pm}/K^{\pm}&\rightarrow
&\mu^{\pm}+\nu_{\mu}(\bar{\nu}_{\mu})\\
& & & &\mu^{\pm}\rightarrow e^{\pm}+\nu_{e}(\bar{\nu}_{e})+
\bar{\nu}_{\mu}(\nu_{\mu}).
\end{array}
\label{atmo-nu-prod}
\end{equation}
Thus, when the conditions for the decay of all these particles are satisfied, we expect flavor ratios for the atmospheric neutrino fluxes of $\nu_\mu/\bar{\nu}_\mu\sim 1$ and $(\nu_\mu+\bar{\nu}_\mu)/(\nu_e + \bar{\nu}_e) \sim 2$.  This flux is usually referred to as the conventional atmospheric neutrino flux and the energies of these atmospheric neutrinos range from a few MeV up to hundreds of TeV.

The gross features of the spectrum are easy to understand (see e.g., 
Refs.~\cite{Gaisser:1990, Gaisser:2002jj}).  For energies
$E_\nu~\ll~100~\textrm{GeV}/|\cos\theta|$, with $\theta$ being the
zenith angle, all pions decay and the neutrino spectrum has
approximately the same power as the primary cosmic-ray spectrum, which
is $\sim E^{-2.7}$.  Above these energies, the expected power-law
neutrino spectrum (from pion decays) is asymptotically one power
steeper.  This has to do with the extra factor of $1/E_\pi$ in the
ratio of the pion decay length to interaction length.  The fact that the flux is higher in the horizontal direction (it is proportional to
$\sec\theta$) is explained by the longer decay path near the horizon: a
pion traveling through the atmosphere horizontally ($\cos\theta \simeq
0$) has a higher probability of decaying in the atmosphere than a pion
traversing the atmosphere vertically ($\cos\theta \simeq 1$) and
therefore the flux is higher for the horizontal component.  A similar
behavior occurs for kaons, but shifted to energies about one order of
magnitude higher.  On the other hand, the muon spectrum is very
similar to the parent meson spectrum, but for energies above
$\sim$~1~GeV, many muons reach the ground before decaying (at
$\sim$~100~GeV, only $\sim$~15\% of atmospheric muon neutrinos come
from muon decay), and hence the neutrino flavor ratio grows with
increasing energy.  Similarly to the angular dependence of neutrinos
from pion and kaon decays, this results in the steepening of the
electron neutrino spectrum occurring at lower energies for vertical
directions and at higher energies for propagation close to the
horizon.  Existing analytical approximations to the atmospheric
neutrino fluxes versus energy and zenith angle at high energies also
allow a good qualitative understanding of all these
features~\cite{Gaisser:1990, Gaisser:2002jj}. 

The huge range of energies and baselines provided by the atmospheric
neutrinos opens up the possibility of exploring many different and
exciting physics topics, from the measurement of neutrino mixing
parameters~\cite{Wendell:2010md, Abe:2012jj} to more exotic
ones~\cite{Liu:1997yb, Fornengo:2001pm, Beacom:2003zu,
  GonzalezGarcia:2004wg, Friedland:2004ah, Ando:2007ds,
  Choubey:2007ji, Abe:2008zza, GonzalezGarcia:2008ru, Abbasi:2009nfa,
  Gandhi:2011jg, Mitsuka:2011ty, Asaka:2012hc, Esmaili:2012nz}.  At
very high energies, above 10~TeV, neutrino interaction cross sections
become high enough for neutrinos going through the Earth to start
becoming attenuated.  This effect is sensitive to the neutrino
interaction cross sections and to the density profile of the Earth and
thus, a measurement of the atmospheric neutrino-induced event rate at
these energies could provide information about the Earth's internal
structure~\cite{GonzalezGarcia:2007gg, Borriello:2009ad,
  Takeuchi:2010, Romero:2011zzb}.  The intermediate-energy region,
between $\sim$~50~GeV and $\sim$~1~TeV, provides information about the
atmospheric neutrino flux normalization~\cite{Abbasi:2009nfa}.
Finally, in the low-energy region which we exploit here (below about
50~GeV), neutrino oscillation effects could be
significant~\cite{Banuls:2001zn, Petcov:1998su, Akhmedov:1998ui, Petcov:1998, Akhmedov:1998xq, Chizhov:1998ug, Chizhov:1999az, Chizhov:1999he, Akhmedov:2005yj, Akhmedov:2006hb, Bernabeu:2001xn, Bernabeu:2003yp, Petcov:2004bd}, in particular due to the large value of the $\tet$ mixing angle recently measured~\cite{An:2012eh, Ahn:2012nd, Abe:2012tg, Abe:2011sj}.  Resonant matter effects in the
propagation of neutrinos inside the Earth could be very important for
$E_\nu \sim [1, \, 15]$~GeV and could strongly enhance or decrease the
oscillation probabilities, for neutrinos in the case of normal
neutrino mass hierarchy and for antineutrinos if the neutrino mass
hierarchy is inverted. 

The atmospheric neutrino fluxes we use in this work are those from
Ref.~\cite{Barr:2004br} (see also Refs.~\cite{Battistoni:2002ew,
Honda:2011nf, Athar:2012it} for other atmospheric neutrino flux
calculations\footnote{Note that in Ref.~\cite{Athar:2012it} the atmospheric neutrino flux at the South Pole was calculated.  At this site there is no rigidity cutoff, and the expected flux is slightly higher than that at the Kamioka site where there is an intermediate rigidity cutoff.  However, the main differences show up at low energies ($E_\nu \ll 10$~GeV) and are not expected to significantly affect our results.}).  The absolute electron and muon atmospheric (anti)neutrino fluxes are uncertain at the level of 10\%~-~20\% in the energy region of interest here, which is related to our ignorance regarding hadron production models.  The uncertainties quoted above are largely canceled for the muon neutrino-antineutrino flavor ratio as well as for the up-down ratio, the former expected to be of order $1\%$ above 1~GeV~\cite{Barr:2006it}.  Accelerator data from the HARP, MIPP and NA61 experiments on particle multi-production are also expected to improve the predictions of the absolute atmospheric neutrino fluxes~\cite{Schroeter:2010zz}.  However, as we will see when discussing the effects of correlated systematic uncertainties in the analysis, the uncertainties related to the normalization of the atmospheric fluxes are alleviated when higher energy bins, as well as different angular bins, in which oscillation effects are negligible, are also used.

\subsection{The Earth's internal structure}

Most of the knowledge we have about the internal structure of the
Earth and the physical properties of its different layers comes from
geophysics and, in particular, from the data we obtain from seismic
waves. Other information, from geomagnetic and geodynamical
data, solid state theory and high temperature-pressure experimental
results is also used.  It turns out that a reliable estimate of
the density of the Earth is essential to solve a number of important
problems in geophysics, such as the dynamics of the core and mantle, the gravity field, the mechanism of the geomagnetic dynamo, the bulk
composition of the Earth, etc. (see, e.g., Ref.~\cite{Bolt:1991} and
references therein).

The Earth is conventionally divided into three main (approximately)
spherical concentric shells: the crust, the mantle and the core, each
of them further divided into subshells, with different
properties~(see, e.g., Refs.~\cite{Loper:1995, Anderson:2007}).
Whereas the crust only represents about 0.4\% of the Earth's mass, the
mantle and the core constitute about 68\% and 32\%, respectively.  On
the other hand, the core has a radius almost half that of the Earth
and it is about twice as dense as the mantle. 

The divisions of the Earth's interior are based on the reflection and
refraction of compressional and shear waves, i.e., P- and S-waves,
respectively. The crust is the brittle outer shell, with a large
proportion of incompatible elements.  It is composed of three
different crustal types: continental (thick and composed primarily of
granite), transitional (defined by the islands, island arcs and
continental margins) and oceanic (thinner and composed primarily of
basalt).  The crust is separated from the mantle by a boundary
discovered by Mohorovicic in 1909.  The mantle is composed mainly of
silicates and oxides, being a poor conductor of electricity and heat
and with very high temperature-dependent viscosity.  Its temperature
increases with depth creating a geothermal gradient which is
responsible for the different rock behaviors that determine the mantle
subdivisions.  Whereas in the upper mantle rocks are cool and brittle,
in the lower mantle they are hot and soft, but not molten.  In 1891
E.~Wiechert inferred the existence of a completely different region
beneath the mantle from the mean density and moment of inertia of the
Earth, but it was not until 1914 that B.~Gutenberg made the first
accurate determination of the location of the core-mantle
boundary~\cite{Gutenberg:1914} from the R.~D.~Oldham measurement in
1906 of a sharp decrease in the velocities of P- and S-waves deep in
the Earth.  Soon after this, the first models of the radial
structure of elastic velocities started to be developed.  This
boundary presents a density contrast of a factor of $\sim$~2 and a
viscosity contrast of $\sim$~20-24 orders of magnitude, which is what
ultimately causes the strongly differing time and length scales for
convection motions in these two layers.  The Earth's core, the Earth's
source of internal heat, is thought to be composed mainly of an alloy
of iron and nickel, with a small admixture of lighter elements, which
is a good conductor of electricity and heat.  Its radius is
approximately 2900~km and it is divided into the outer (liquid) and
the inner (solid) cores.

\begin{figure}[tp]
\centering
\includegraphics[width=0.6\textwidth]{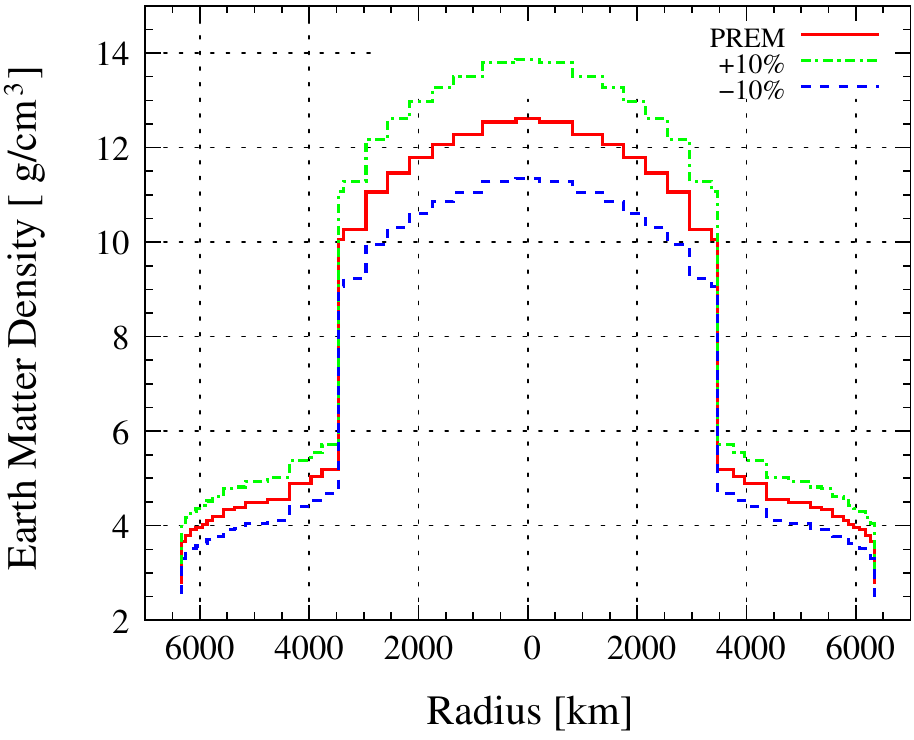}
\mycaption{\label{fig:prem} {\sl \textbf{The Earth's density profile} according to the PREM~\cite{Dziewonski:1981xy} (red) and with a $\pm 10\%$ overall density fluctuation (green dash-dotted and blue dashed  lines), as a function of the distance from the center of the Earth.}}   
\end{figure}

Although the Earth is not exactly spherically symmetric and is
irregular, the use of mean models, taking the Earth's physical
parameters as symmetric, is very useful as a reference
framework for different studies~\cite{Khan:1983}.  During many years,
the most widely used radially symmetric model has been the Preliminary
Reference Earth Model (PREM)~\cite{Dziewonski:1981xy}, which included
anisotropy in the upper mantle, and it was a good fit to free
oscillation center frequency measurements, surface-wave dispersion
observations, travel-time data for body-waves and some astronomical
data such as the Earth's radius, mass and moment of inertia.  However,
there was no discussion about the sources of error or covariances.  A
more recent 1-D model, AK135, which combines travel times and free
oscillations, is also available~\cite{Kennett:1995, Montagner:1995}.
Although there are observed deviations at the few per cent level that
cannot be explained by 1-D models, to first order, the structure of
the Earth is determined by its radial properties and these reference
models are used as a starting point for more refined studies.  In
Fig.~\ref{fig:prem} we show the matter density profile of the Earth
(and fluctuations of $\pm 10\%$) as a function of the distance to the
center according to the PREM~\cite{Dziewonski:1981xy}.  Up to a radius
of $\sim 2900$~km, the density increases gradually from the crust to
the mantle. Then, in the core--mantle boundary, there is a very sharp
transition, followed by a gradually increasing density until the
center of the Earth.

For the last half a century the problem of determining the density
distribution of the Earth has mainly consisted of perturbation
inversion using different seismic data.  This is a very demanding
problem, which in many cases is non-linear.  Moreover, most studies
of the Earth's radial structure make inferences about different
properties given a specific density model or based on empirical
relations between seismic waves velocities and density.  Nevertheless,
there are significant trade-offs with other crucial parameters because wave velocities also depend on composition, temperature, pressure and elastic properties, which necessarily imply some modeling and uncertainties in the density estimate.  Recently, 3-D models, using
1-D models as a reference, have been developed, although some studies
concluded that few robust density features could be constrained with
existing data~\cite{Resovsky:1999, Masters:2000, Romanowicz:2001, Kuo:2002} (see, however, Refs.~\cite{Ishii:1999, Ishii:2001, Ishii:2002, Resovsky:2002, Ishii:2004, Hafner:2012}).  All in all,
density values, averaged over $\sim$~100~km are thought to be known at
the level of a few per cent at all depths, whereas density gradients
are less known.  On the other hand, linear integral constraints on the
spherically symmetric density distribution are known at the level of
$\sim 10^{-4}$ from estimates of the Earth's mass and moment of
inertia~\cite{Chambat:2001, Luzum:2011, AAlmanac} (see also
Ref.~\cite{Hafner:2012} for a recent analysis of the spectra of the
lowest frequency seismic mode with a precision of $\sim 10^{-3}$).
This implies that global variations of the density are constrained to
be within those uncertainties.  Obviously then, overall density
variations at the level of a few per cent (as those shown in
Fig.~\ref{fig:prem}) are already excluded.  Nevertheless, we will not
impose these linear integral constraints and evaluate, using the PREM
as our reference model, how sensitive atmospheric neutrino oscillation
tomography is to changes in the density.

\subsection{Oscillation probabilities}
\label{subsec:probabilities}

For neutrino energies in the range of a few GeV, the transition probabilities $\nu_{\mu} \rightarrow \nu_{e}$ ($\bar{\nu}_{\mu} \rightarrow \bar{\nu}_{e}$) and $\nu_{e} \rightarrow \nu_{\mu (\tau)}$
($\bar{\nu}_{e} \rightarrow \bar{\nu}_{\mu (\tau)}$) of atmospheric
neutrinos in their propagation through the Earth are relevant if
genuine 3-flavor neutrino mixing takes place, i.e., if the $\tet$ mixing angle is different from zero~\cite{Banuls:2001zn, Bernabeu:2001jb, Bernabeu:2002tj, Petcov:1998su, Akhmedov:1998ui, Petcov:1998, Akhmedov:1998xq, Chizhov:1998ug, Chizhov:1999az, Chizhov:1999he, Akhmedov:2005yj, Akhmedov:2006hb}.  Moreover, in this energy range and for these baselines ($L > 1000$~km), CP-violation effects are very small and can be safely neglected.  Likewise, effects due to the 1-2 sector are also subdominant and, as a first approximation, can also be neglected.  In this context, the calculation of the transition probabilities effectively reduces to a 2-neutrino problem, with $\ma$ and $\tet$ playing the role of the relevant 2-neutrino oscillation parameters.  Within these approximations, the 3-neutrino oscillation probabilities of interest for atmospheric $\nu_{e,\mu}$ having energy $E_\nu$ and crossing the Earth along a trajectory characterized by a zenith angle\footnote{Neutrinos with $\cos\theta = - 1$ are directly upgoing and traverse the entire diameter of the Earth, those with $\cos\theta = 0$ come from the horizon and those with $\cos\theta > 0$ are downgoing and reach the detector from above.} $\theta$, have the following form~\cite{Petcov:1998su, Chizhov:1998ug, Chizhov:1999az,
Chizhov:1999he} (see also Refs.~\cite{Akhmedov:1998ui, Akhmedov:1998xq, Akhmedov:2005yj, Akhmedov:2006hb}):  
\ba
P_{3\nu}(\nu_e \rightarrow \nu_e) & \simeq & 1 - P_{2\nu} \, , 
\label{eq:P3ee} \\
P_{3\nu}(\nu_e \rightarrow \nu_\mu) & \simeq & P_{3\nu}(\nu_\mu
\rightarrow \nu_e) \simeq \sa \, P_{2\nu} \, , 
\label{eq:P3emu} \\
P_{3\nu}(\nu_e \rightarrow \nu_\tau) & \simeq & \cos^2\tmt \, P_{2\nu} \, , \label{eq:P3etau} \\
P_{3\nu}(\nu_\mu \rightarrow \nu_\mu) & \simeq & 1 - \frac{1}{2} \,
\sta - \sin^4\tmt \, P_{2\nu} + \frac{1}{2} \, \sta \, Re~( e^{-i\kappa} A_{2\nu}(\nu_\tau \rightarrow \nu_\tau)) \, , 
\label{eq:P3mumu} \\
P_{3\nu}(\nu_\mu \rightarrow \nu_\tau) & = & 1 - P_{3\nu}(\nu_\mu
\rightarrow \nu_\mu) - P_{3\nu}(\nu_\mu \rightarrow \nu_e) \,  
\label{eq:P3mutau} 
\ea
where $P_{2\nu} \equiv P_{2\nu}(\ma, \tet ; E_\nu, \theta)$ is the 2-neutrino probability describing $\nu_e \rightarrow \nu_x$ transitions, where $\nu_x = \sin\tmt \, \nu_\mu + \cos\tmt \, \nu_\tau$, and $\kappa$ and $A_{2\nu}(\nu_{\tau} \rightarrow \nu_{\tau}) \equiv A_{2\nu}$ are the phase and the 2-neutrino transition probability amplitude.  For antineutrinos the oscillation probabilities are analogous to those for neutrinos: they can be obtained formally from Eqs.~(\ref{eq:P3ee}) - (\ref{eq:P3mutau}) by changing the sign of the matter potential (or equivalently, $\rho$ by $-\rho$).  It is interesting to note that, within the approximation $\ms = 0$, the probabilities for neutrinos and NH (IH) are the same as those for antineutrinos and IH (NH).

Therefore, the magnitude of the matter effects depends on the
2-neutrino oscillation probability $P_{2\nu}$.  In case of oscillations in vacuum, $P_{2\nu} \sim \stch$, so this probability is small.  However, matter effects can strongly enhance $P_{2\nu}$ and thereby greatly modify the 3-neutrino probabilities.  On the other hand, if $\tet = 0$, then $P_{2\nu} = 0$ and $Re~(e^{-i\kappa} A_{2\nu}(\nu_\tau \rightarrow \nu_\tau)) = \cos\left(\ma L/(2 \, E_\nu)\right)$, and hence matter effects are absent.  If this were the situation, these probabilities would get reduced to the case of $\nu_\mu \leftrightarrow \nu_\tau$ 2-neutrino oscillations, so for NH and IH they would be equal and identical to the case of vacuum oscillations.

\begin{figure}[tp]
\includegraphics[width=0.49\textwidth]{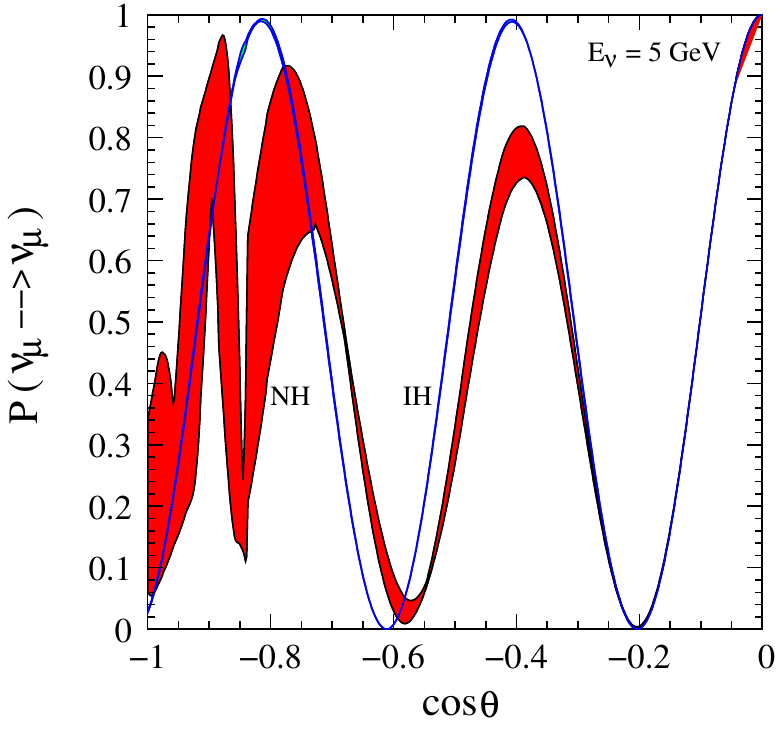}
\includegraphics[width=0.49\textwidth]{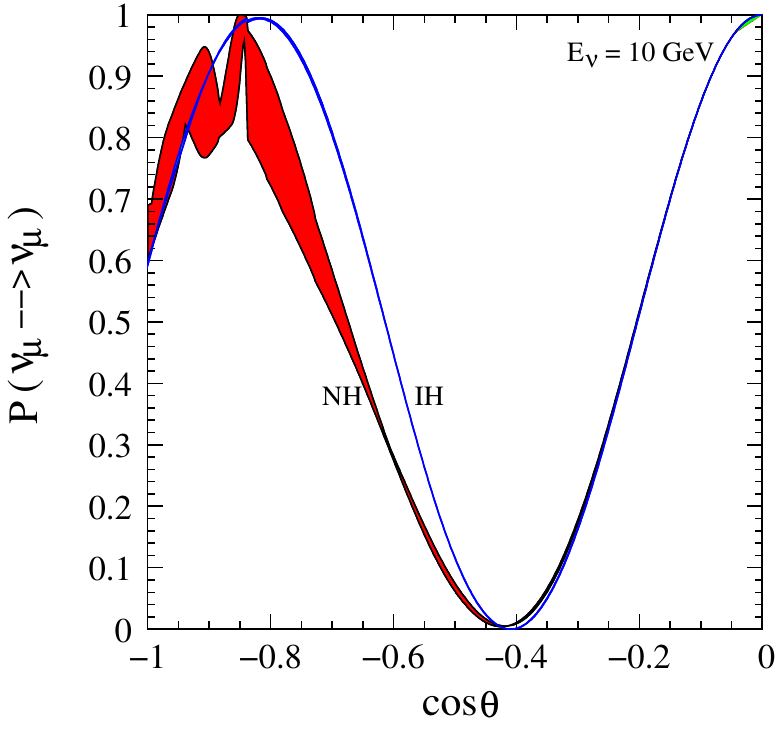}
\vskip0.5cm
\includegraphics[width=0.49\textwidth]{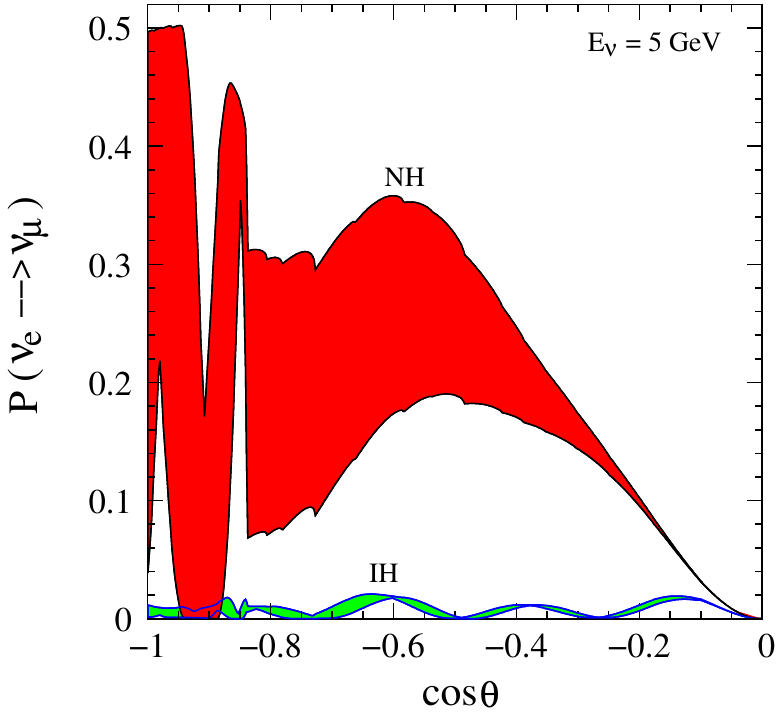}
\includegraphics[width=0.49\textwidth]{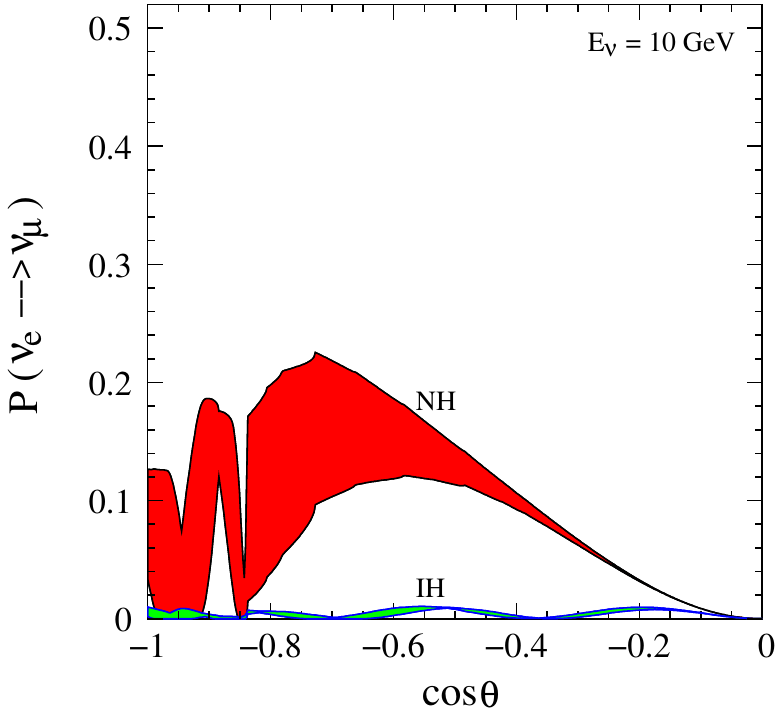}
\mycaption{\label{fig:probability} {\sl \textbf{Oscillation probabilities.} Upper panels: $P(\nu_\mu \rightarrow \nu_\mu)$ as a function of $\cos\theta$, for $E_\nu=5$~GeV (left panel) and $E_\nu=10$~GeV (right panel), for NH (red regions) and IH (green regions limited by blue lines).  Lower panels: Same but for $P(\nu_e \rightarrow \nu_\mu)$.  The widths of the bands correspond to varying the matter density by $\pm 10\%$.  Similar results are obtained for the case of antineutrinos by exchanging the curves for NH (IH) by those for IH (NH).  We have used the current best-fit values for the oscillation parameters~\cite{Tortola:2012te}.}
}
\end{figure}

In Fig.~\ref{fig:probability} we show the $\nu_\mu \rightarrow \nu_\mu$
(upper panels) and $\nu_e \rightarrow \nu_\mu$ (lower panels)
transition probabilities as a function of $\cos\theta$ for two
different energies ($E_\nu=5$~GeV in left panels and $E_\nu=10$~GeV in
right panels) for NH (red regions) and IH (green regions limited by blue lines).  The bands correspond to how the probabilities change if the density (according to the PREM) varies up to $\pm 10\%$.  We can see that matter effects, that are very sensitive to the value of the density,  tend to greatly enhance the $\nu_e \leftrightarrow \nu_\mu$ transitions and reduce the $\nu_\mu \rightarrow \nu_\mu$ survival probability, and also shift the positions of the maxima and minima with respect to the case of negligible matter effects.  Moreover, a change in the matter density profile would shift the location of both the resonance energy and the baseline at which the resonant behavior is maximized, modifying the number of events in the different angular and energy bins.  For propagation in the mantle only, the resonance is, to a good approximation, the MSW resonance.  However, in the case of propagation through the core ($\cos\theta < - 0.83$), non-trivial resonant effects show up, although for the energies considered in Fig.~\ref{fig:probability} they are not maximal.  Notice that matter
effects are larger for $E_\nu = 5$~GeV and more baselines are affected than for $E_\nu = 10$~GeV, the reason being that resonances are closer to $E_\nu = 5$~GeV (see below).

As we discussed above, the Earth is composed, in addition to the
crust, of two major density structures that constitute almost all the
mass of the planet: the mantle and the core.  Although in our
calculations we compute the full 3-neutrino evolution and use the
detailed PREM for the density distribution\footnote{Moreover, throughout this work we will consider as a good approximation the Earth to be an isoscalar neutral medium, i.e., $n_e/(n_p+n_n) = 1/2$, where $n_e, n_p, n_n$ are the electron, proton and neutron densities, respectively.}, as a first approximation and in order to gain insight about the physics of neutrino propagation in the Earth, one can consider a simplified model for the density profile, divided into two shells with different densities, $\bar\rho_m \simeq 4.5~\textrm{g/cm}^3$ and $\bar\rho_c \simeq 11.5~\textrm{g/cm}^3$,
with the core-mantle boundary at a radius of $\sim$~2900~km.  All the
interesting features of the atmospheric neutrino oscillations in the
Earth can be understood quite accurately within this
framework~\cite{Petcov:1998su, Akhmedov:1998ui, Petcov:1998, Akhmedov:1998xq, Chizhov:1998ug, Chizhov:1999az, Chizhov:1999he, Akhmedov:2005yj, Akhmedov:2006hb}.  There are analytical solutions for the transition probabilities for neutrinos crossing the Earth~\cite{Petcov:1998su, Akhmedov:1998ui, Petcov:1998, Akhmedov:1998xq, Chizhov:1998ug, Chizhov:1999az, Chizhov:1999he, Akhmedov:2005yj, Akhmedov:2006hb, Bernabeu:2001xn, Bernabeu:2003yp}, but they reduce to the case of neutrino propagation in a medium of constant density for trajectories such that $\cos\theta > - 0.83$, i.e., for neutrinos which propagate only in the mantle and could experience MSW resonant effects~\cite{Banuls:2001zn}.  In this case, Eqs.~(\ref{eq:P3ee}) - (\ref{eq:P3mumu}) read\footnote{Note that for a spherically symmetric Earth, the neutrino trajectory is fully specified by its zenith angle, $L = - 2 \, R_\oplus \, \cos\theta$.} 
\ba
P_{3\nu}(\nu_e \rightarrow \nu_e) & = & 1 - \sin^22\theta^m_{13} \, 
\sin^2 \left(\frac{\Delta^m_{31} \, L}{2}\right) \, ,   
\label{eq:P3eeC} \\
P_{3\nu}(\nu_e \rightarrow \nu_\mu) & = & \sa \, \sin^22\theta^m_{13} \, \sin^2 \left(\frac{\Delta^m_{31} \, L}{2}\right) \, ,  
\label{eq:P3emuC} \\ 
P_{3\nu}(\nu_e \rightarrow \nu_\tau) & = & \cos^2\tmt \, \sin^22\theta^m_{13} \, \sin^2 \left(\frac{\Delta^m_{31} \, L}{2}\right) \, ,   
\label{eq:P3etauC} \\
P_{3\nu}(\nu_\mu \rightarrow \nu_\mu) & = & 1 - \frac{1}{2}
\sta - \sin^4\tmt \, \sin^22\theta^m_{13} \, \sin^2 \left(\frac{\Delta^m_{31} \, L}{2}\right) +   
\label{eq:P3mumuC} \\
& &  \frac{1}{2} \sta \left[\cos\left(\frac{\Delta_{31}
    \, L}{2} \left(1-\frac{A}{\Delta m_{31}^2}\right)\right)
  \cos\left(\frac{\Delta^m_{31} \, L}{2}\right) -  \right. 
  \nonumber \\ 
& & \left. \cos 2 \theta^m_{13} \sin\left(\frac{\Delta_{31} \, L}{2}
  \left(1-\frac{A}{\Delta m_{31}^2}\right)\right)
  \sin\left(\frac{\Delta^m_{31} \, L}{2}\right)  \right] \, ,
\ea
where
\ba
\Delta_{31} & \equiv & \frac{\ma}{2 \, E_\nu} \, , \\
\Delta^m_{31} & \equiv & \frac{1}{2 \, E_\nu} \, \sqrt{(\ma \, \cos2\tet -A)^2+ (\ma \sin2\tet)^2} \, ,\\ 
\sin^22\tilde{\theta}_{13} & \equiv & \stch \left(\frac{\Delta}{\Delta^m}\right)^2 \, ,  
\ea
and the matter parameter is $A \equiv 2 \, \sqrt{2} \, G_F \, n_e \,
E_\nu$, with $G_F$ the Fermi constant and $n_e$ the electron number
density.  The expressions for antineutrinos can be obtained by
replacing $A \rightarrow -A$.  Thus, the resonant behavior, when
maximal mixing occurs, happens for the case of NH (IH) in the neutrino
(antineutrino) channel at the resonant energy,
\be
E_{\rm res} \equiv \frac{\ma \, \cos2\tet}{2 \, \sqrt{2} \, G_F \, n_e} \simeq 7 \, \textrm{GeV} \, \left(\frac{4.5 \, \textrm{g/cm}^3}{\rho}\right) \, \left(\frac{\ma}{2.4 \times 10^{-3} \, \textrm{eV}^2}\right) \, \cos2\tet \, . 
\ee
On the other hand, the first oscillation maximum of the oscillation
term happens when the oscillation phase is $\frac{\Delta^m \, L}{2}=\frac{\pi}{2}$.  Hence, the baseline at which both the conditions
for the resonance and the first oscillation maximum are satisfied,
is~\cite{Banuls:2001zn} 
\be
L_{\rm max} = \frac{\pi}{\sqrt{2} \, G_F \, n_e \, \tan2\tet} \simeq 1.1 \cdot 10^4 \, \textrm{km} \, \left(\frac{4.5
  \, \textrm{g/cm}^3}{\rho}\right) \, \left(\frac{1/3}{\tan2\tet}\right) \, . 
\ee
%

\begin{figure}[t]
\centering
\includegraphics[width=0.6\textwidth]{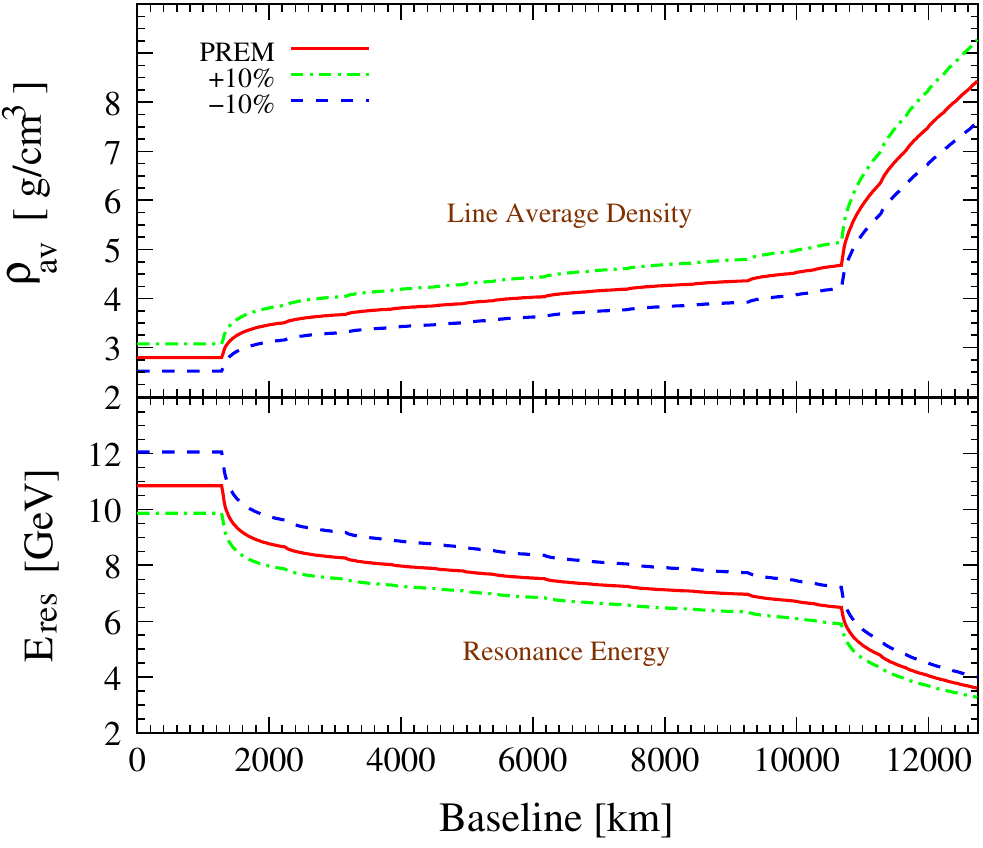}
\mycaption{\label{fig:reso} {\sl \textbf{Line average of the Earth's density and resonance energy.} Upper panel: Line average density of the Earth for a given trajectory with total baseline $L$.  Lower panel: Resonance energy for propagation of neutrinos in the Earth assuming the density is constant and equal to the line average density.  We have assumed $\sch = 0.025$, $\ma = 2.44 \times 10^{-3}$~eV$^2$ and the density distribution according to the PREM (red lines) and with a $\pm 10\%$ overall density fluctuation (green dash-dotted and blue dashed  lines).}
}  
\end{figure}

\noindent
For $\stch \sim 0.1$, $L_{max}\simeq 1.1 \cdot 10^4$~km and $E_{\rm res}\simeq 7$~GeV for densities approximately corresponding to the average value of the mantle.  However, strong matter effects are also present around the resonant energy even if the baseline does not correspond exactly to the first oscillation maximum, i.e., even for baselines $L \simeq L_{\textrm{max}}/\sqrt{2}$~\cite{Banuls:2001zn}.  For illustration purposes, we show in Fig.~\ref{fig:reso} the line average density for a given trajectory in the Earth (upper panel) and the corresponding resonance energy in the case of propagation in a medium of constant density with that value (lower panel). 

Using Eqs.~(\ref{eq:P3ee}) - (\ref{eq:P3mutau}) it is straightforward to see that the fluxes of atmospheric $\nu_{e,\mu}$ of energy $E_\nu$ which reach the detector after crossing the Earth along a given trajectory, $\Phi_{\nu_{e,\mu}}(E_\nu,\theta)$, are given by the following expressions in the case of the 3-neutrino oscillations under discussion~\cite{Petcov:1998, Akhmedov:1998xq, Chizhov:1998ug, Bernabeu:2003yp}: 
\ba
\Phi_{\nu_e}(E_\nu, \theta) & \simeq & \Phi^{0}_{\nu_e} \, \left[1 + (\sa \, r - 1) \, P_{2\nu}\right] \, ,
\label{eq:Phie} \\
\Phi_{\nu_{\mu}}(E_\nu, \theta) & \simeq & \Phi^{0}_{\nu_{\mu}}
\left[1 - \frac{1}{2} \sta \, + \right. \nonumber \\  
& & 
\left. \sin^4\tmt \, \left( (\sa \, r)^{-1} - 1\right) \, P_{2\nu} + \frac{1}{2} \sta \, Re~( e^{-i\kappa} A_{2\nu}(\nu_{\tau} \rightarrow \nu_{\tau}))\right] \, ,  
\label{eq:Phimu} 
\ea
where $\Phi^{0}_{\nu_{e(\mu)}} = \Phi^{0}_{\nu_{e(\mu)}}(E_\nu, \theta)$ is the $\nu_{e(\mu)}$ flux at the production point in the atmosphere and
\be
r \equiv r(E_\nu, \theta) \equiv \frac{\Phi^{0}_{\nu_{\mu}}(E_\nu,
  \theta)}{\Phi^{0}_{\nu_{e}}(E_\nu, \theta)} \, . 
\label{eq:r}
\ee
For neutrino trajectories with $\cos\theta < - 0.4$, i.e., $L >
5100$~km, the predicted ratio of $\nu_\mu$ to $\nu_e$ atmospheric
neutrino fluxes is $r \simeq 2.0-2.2 \, \, (2.1-2.6)$ for sub-GeV
(anti)neutrinos while $r \simeq 2.1-5.6 \, \, (2.4-7.2)$ for
multi-GeV (anti)neutrinos~\cite{Barr:2004br, Battistoni:2002ew, Honda:2011nf, Athar:2012it}.  Thus, even in the case of resonant matter effects taking place, the effects of $\tet$-driven transitions are suppressed for the sub-GeV sample due to the term $(\sa \, r - 1)$, especially if $\sa \lesssim 0.5$ (the current $3\sigma$ confidence level (CL) allowed range is $\sa = (0.36 - 0.68)$~\cite{Tortola:2012te}).  On the other hand, for multi-GeV neutrinos, $(\sa \, r - 1) > 0$, so actually this factor can amplify the matter effects.  Nevertheless, neutrino telescopes are \v{C}erenkov detectors, so they have no charge-identification capabilities and therefore cannot distinguish neutrinos from antineutrinos.  As the resonances take place for NH for neutrinos and IH for antineutrinos, the $\tet$-driven matter effects would consequently be smeared and the sensitivity to them reduced as compared to the case when measurements of the neutrino-induced rates and antineutrino-induced rates can be performed separately\footnote{The distinction between the neutrino and antineutrino channels can be achieved with magnetized detectors such as the magnetized iron calorimeter at the India-based Neutrino Observatory (INO)~\cite{Athar:2006yb, Mondal:2012fn}, but with much smaller volumes than neutrino telescopes.}.  However, notice that the fact that neutrino and antineutrino cross sections are different could allow us to distinguish NH from IH.  Unfortunately, neutrino telescopes have very poor angular resolution for cascades, so the electron neutrino-induced event rates, where these resonant effects are expected to be larger (see Fig.~\ref{fig:probability}), would give little information.  Therefore, muon tracks are the main signal in this study and hence, the muon neutrino-induced event rates.

Following the discussion above, from Eqs.~(\ref{eq:P3ee}) -
(\ref{eq:P3mumu}), Eqs.~(\ref{eq:Phie}) - (\ref{eq:Phimu}) and the
analogous equations for antineutrinos, we see that $\tet$-driven matter effects increase the $\nu_\mu \leftrightarrow \nu_e$ (and $\bar\nu_\mu \leftrightarrow \bar\nu_e$) transitions with $\sa$, leading to an increase of the electron neutrino-induced event rate and a decrease of the muon neutrino-induced event rate at the detector. The rates are larger for multi-GeV events than for sub-GeV events.

\section{Analysis}
\label{sec:analysis}

\subsection{Detectors set-up}

The IceCube/DeepCore detector~\cite{Collaboration:2011ym} is a
densely instrumented region located at the bottom center of the
IceCube detector at a depth of between 2100~m and 2450~m, in such a way
that it avoids a horizontal layer with a high content of dust and therefore poor optical properties.  The detector consists of six additional strings instrumented with phototubes of higher quantum efficiency with respect to IceCube.  Thus, in general, DeepCore has multiple advantages when compared to Icecube: at the depth at which DeepCore is located, ice is on average twice as clear as the ice above, allowing for a lower neutrino energy threshold.  Additionally, the larger amount of photosensors, separated by 7~m instead of 17~m for IceCube, and the higher quantum efficiency, lead to a significant gain in sensitivity of up to a factor of 6, especially for low energy neutrinos.  The rest of the IceCube detector, along with a horizontal region with additional instrumentation at a depth of 1800~m, could be used as an active veto for downgoing atmospheric muons, allowing the study also of downgoing atmospheric neutrinos.

At the energies of interest here, the angular reconstruction
capabilities of DeepCore for showers is not optimal.  Therefore, for
the present study, we only consider upgoing muon-like events with the
effective volume for the 86-string configuration (IC86) at trigger
(SMT3) and online filter level in the 10-15~GeV energy region as shown in Tab.~\ref{tab:detectors}~\cite{Koskinen}.  We follow a very
conservative approach by only considering this single energy bin
($E_\nu = [10, \, 15]$~GeV) and discarding lower energy bins which in principle have a lower effective volume and worse detection capabilities.  The IceCube Collaboration aims to achieve a signal
efficiency of $\sim 50\%$ for contained and partially contained
events~\cite{Collaboration:2011ym}.

There is also a recent proposal to further upgrade the IceCube detector, with the planned PINGU~\cite{Koskinen:2011zz}.  It would consist of 20 additional strings within the DeepCore volume to lower the neutrino energy threshold down to $\mathcal{O}(1)$~GeV energies and the detector would also benefit from the DeepCore strings.

In this work, we consider two possible PINGU scenarios: a conservative scenario which we refer to as PINGU-0 with 5~GeV energy threshold and with two 5~GeV energy bins ($E_\nu = [5, \, 10]$~GeV and $[10, \, 15]$~GeV), and a less conservative scenario which we call PINGU-I, with four 2.5~GeV energy bins ($E_\nu = [5.0, \, 7.5]$~GeV, $[7.5, \, 10.0]$~GeV, $[10.0, \, 12.5]$~GeV and $[12.5, \, 15.0]$~GeV).  The effective masses for these configurations (Tab.~\ref{tab:detectors}) are computed assuming a trigger setting of 3 digital optical modules hit in 2.5~$\mu s$ that are in local coincidence, a containment criterium which is implemented by a cut on the $z$-position that matches the DeepCore fiducial volume ($-500~\textrm{m} < z < -157$~m) and a tight radius cut from string 36 ($r < 150$~m) which is the center of DeepCore/PINGU~\cite{Koskinen}.

\begin{table}[tp]
\begin{center}
\begin{tabular}{||c||c||c||} \hline\hline
\multicolumn{1}{||c||}{{\rule[0mm]{0mm}{6mm}{Detector}}} 
& \multicolumn{1}{c||}{\rule[-3mm]{0mm}{6mm}{Energy bins (GeV)}} 
&  \multicolumn{1}{c||}{\rule[-3mm]{0mm}{6mm}{$\rho_{\textrm{ice}} \,
    V_{\rm{eff}}$ (Mton)}} 
\cr
\hline \hline
            & 8.9 - 10.875   & 3.91 \cr
DeepCore    & 10.875 - 12.85 & 4.46 \cr
            & 12.85 - 14.825 & 5.69 \cr
            & 14.825 - 16.8  & 6.53 \cr
\hline
            & 5  - 6  & 4.21  \cr
            & 6  - 7  & 4.87  \cr
            & 7  - 8  & 4.91  \cr
            & 8  - 9  & 5.79  \cr
PINGU       & 9  - 10 & 7.54  \cr
            & 10 - 11 & 6.73  \cr
            & 11 - 12 & 6.77  \cr
            & 12 - 13 & 7.58  \cr
            & 13 - 14 & 7.86  \cr
            & 14 - 15 & 8.77  \cr
\hline \hline
\end{tabular}
\mycaption{{\sl \textbf{Detectors set-up.} Effective mass for DeepCore and PINGU for the relevant energy bins which are considered in the current study (see text for details)~\cite{Koskinen}.  In all cases, we add a post-trigger detection efficiency of $50\%$ (not included in these numbers).  For DeepCore we consider a single energy bin, $E_\nu = [10, \, 15]$~GeV, and for PINGU we consider events in the interval $E_\nu = [5, \, 15]$~GeV but we use two different sets of energy bins: for PINGU-0, $E_\nu = [5, \, 10]$~GeV and $E_\nu = [10, \, 15]$~GeV; and for PINGU-I, $E_\nu = [5, \, 7.5]$~GeV, $E_\nu = [7.5, \, 10]$~GeV, $E_\nu = [10, \, 12.5]$~GeV and $E_\nu = [12.5, \, 15]$~GeV.  The size of the angular bins is always $0.1$ in $\cos\theta$ and we
consider events in the interval $\cos\theta = [-1, \, 0]$.}
} 
\label{tab:detectors}
\end{center}
\end{table}

Throughout this work we assume that the determination of the neutrino energy and direction could be achieved with some degree of accuracy.  This implies that in addition to the characterization of the muon tracks induced by the atmospheric muon neutrino interactions in the detector, the accompanying hadronic cascades must also be detected.  This is a challenging task, so in this respect, the results presented here should be taken as optimistic.  Nevertheless, we have assumed very wide energy bins, so that a neutrino energy resolution at the level of $\lesssim 20\%$ is not expected to significantly affect our results. On the other hand, if the angular resolution is not better than the typical root mean square value of the scattering angle between the incoming neutrino and the produced muon, $\theta_{\nu_\mu-\mu} \sim \sqrt{\textrm{GeV}/E_\nu}$, some smearing is expected and hence, the sensitivity would be slightly worse.  Let us also note that the effective masses used in this work represent preliminary MonteCarlo results based on the outcome of ongoing simulations by the IceCube Collaboration and these numbers might be further refined in the near future. 

In summary, for the DeepCore detector simulations we assume a single
energy bin of $E_\nu = [10, \, 15]$~GeV. For the PINGU-0 scenario, we
assume two energy bins: $E_\nu = [5, \, 10]$~GeV and $[10, \, 15]$~GeV.  For PINGU-I, we assume four energy bins: $E_\nu = [5.0, \, 7.5]$~GeV, $[7.5, \, 10.0]$~GeV, $[10.0, \, 12.5]$~GeV and $[12.5, \, 15.0]$~GeV.  The effective masses we consider are shown in Tab.~\ref{tab:detectors}.  In all configurations we assume that the post-trigger efficiency of the detector, for all energy bins, is $50\%$ and consider an angular binning in $\cos\theta$ of width 0.1 for $\cos\theta\in[-1,0]$.

\subsection{Number of muon-like events}

\v{C}erenkov detectors do not have charge-identification capabilities, so neutrinos cannot be distinguished from antineutrinos.  Hence, these two types of events must be added together in the analysis.  The number of contained atmospheric muon-like events (neutrino- plus antineutrino-induced samples) at the detector in a given angular ($i$) and energy bin ($j$), is given by
\ba
(N_\mu)_{ij} & = & 2 \pi \, N_A \, T \, 
  \int_{(\cos\theta)_i}^{(\cos\theta)_{i+1}} d\cos\theta
  \int_{(E_\nu)_j}^{(E_\nu)_{j+1}} dE_\nu \, \epsilon \, 
\rho_{\textrm{ice}} \, V_{\textrm{eff}} (E_\nu) \times \nonumber \\
 & & \left[
\left(\frac{d\Phi_{\nu_\mu}}{d\cos\theta \, dE_\nu} \, P(\nu_\mu
  \rightarrow \nu_\mu) + \frac{d\Phi_{\nu_e}}{d\cos\theta \, dE_\nu}
  \, P(\nu_e \rightarrow \nu_\mu)\right) \, \sigma^{\nu
   N}_{\textrm⁄CC}(E_\nu) +  \right. \nonumber \\
 & & \left.
\left(\frac{d\Phi_{\bar\nu_\mu}}{d\cos\theta \, dE_\nu} \, P(\bar\nu_\mu
  \rightarrow \bar\nu_\mu) + \frac{d\Phi_{\bar\nu_e}}{d\cos\theta \,
    dE_\nu} \, P(\bar\nu_e \rightarrow \bar\nu_\mu)\right) \,
  \sigma^{\bar\nu N}_{\textrm⁄CC}(E_\nu) \right] \, , 
\ea  
where $N_A$ is the Avogadro number, $T$ is the time of data taking, $\epsilon = 0.5$ is the post-trigger efficiency, $\rho_{\textrm{ice}}$ is the density of ice, $V_{\textrm{eff}}$ is the effective volume (the effective mass, $\rho_{\textrm{ice}} \, V_{\textrm{eff}}$, is given in Tab.~\ref{tab:detectors} for the different configurations we consider) and $\sigma^{\nu N}_{\textrm⁄CC}$ and $\sigma^{\bar\nu N}_{\textrm⁄CC}$ are the deep inelastic scattering cross sections for neutrinos and antineutrinos off an isoscalar target\footnote{Note that ice is not an isoscalar target.  However, the differences in the cross sections with respect to an isoscalar target are at the level of 2-3\%.  Notice also that the values we use are slightly different from the world-averaged values over energies that extend well above 100~GeV~\cite{Kuzmin:2005bm, Beringer:1900zz, Formaggio:2012}.}, 
\ba
\sigma^{\nu N}_{\textrm⁄CC} & \simeq & 7.30 \cdot 10^{-39} \,
\left(E_\nu/\textrm{GeV}\right) \, \textrm{cm}^2~; \\ 
\sigma^{\bar\nu N}_{\textrm⁄CC} & \simeq & 3.77 \cdot 10^{-39} \,
\left(E_\nu/\textrm{GeV}\right) \, \textrm{cm}^2 \, . 
\ea

\begin{figure}[t]
\centering
\includegraphics[width=0.49\textwidth]{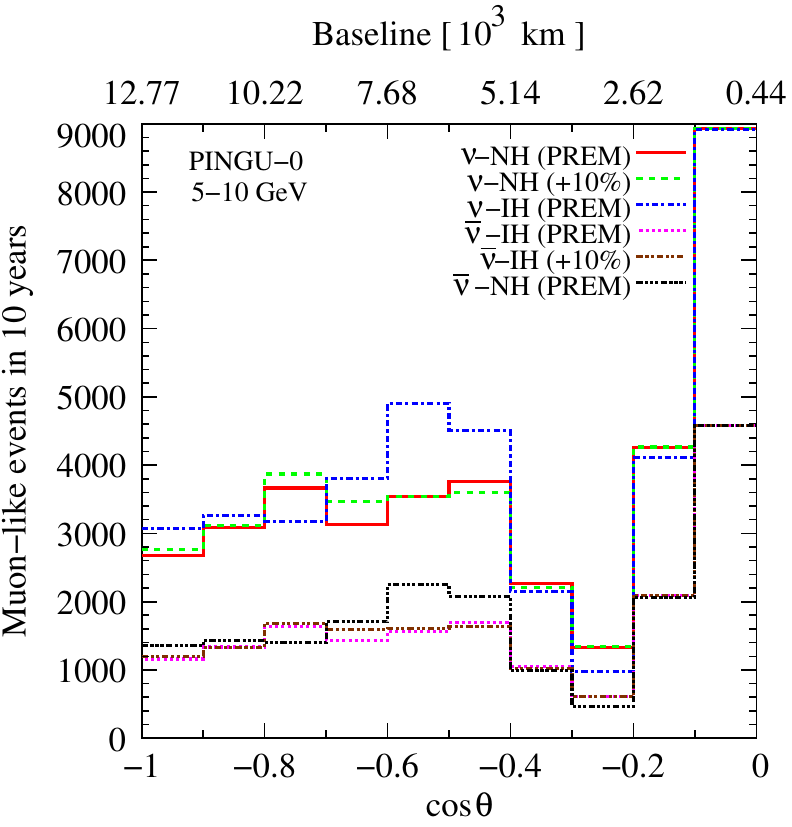}
\includegraphics[width=0.49\textwidth]{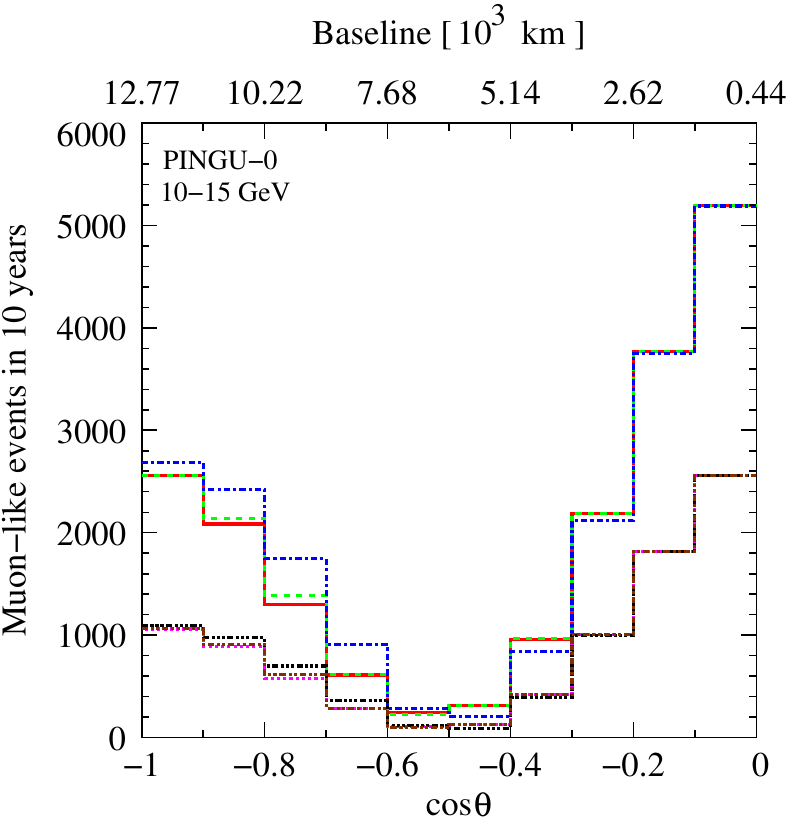}
\mycaption{{\sl \textbf{Number of muon-like contained events in PINGU-0 after 10 years for NH and IH} in the case of a PREM matter profile, in the 5-10~GeV bin (left) and 10-15~GeV (right) as a function of $\cos\theta$ (also shown is the corresponding baseline on the upper axis).  We also show the event numbers in the case of a PREM matter profile with an overall fluctuation of $+10\%$ for neutrino with NH and antineutrino with IH.  Note that the vertical axis is different in each panel.  Although shown independently, in the analyses we add together the neutrino and antineutrino events, due to the lack of charge-identification of the detectors.}
}   
\label{fig:events} 
\end{figure}

The number of atmospheric neutrino- and antineutrino-induced contained
events that are expected in the PINGU-0 set-up after an exposure of 10~years is shown in Fig.~\ref{fig:events} as a function of $\cos\theta$.  The left panel corresponds to the 5-10~GeV bin and the right panel to the 10-15~GeV energy bin.  The number of events for both neutrino mass hierarchies, NH and IH, are shown for neutrinos and antineutrinos.  We also depict two cases where we increase the Earth's density by 10\% with respect to the PREM: neutrino-induced events for NH and antineutrino-induced events for IH, where resonant matter effects are at play.  We show the neutrino and antineutrino rates independently for the sake of illustrating the matter effect taking place for one case or the other, depending on the mass hierarchy, although in the analysis they are added.   

These plots show the dependence of the numbers of muon-like events on
the mass hierarchy (compare the red solid and the blue dotted lines
for neutrinos with the magenta dotted and the black dotted lines for
antineutrinos) and on changes of the normalization of the matter
profile (compare the red solid with the green dotted lines for
neutrinos and NH and the magenta dotted with the brown dotted lines
for antineutrinos and IH).  We only show the results with a change in
the density with respect to the PREM for the neutrino-NH events and
the antineutrino-IH events because matter effects take place inside
the Earth in the case of NH for antineutrinos and of IH for antineutrinos.  The effect is larger for the neutrino channel mainly
due to the larger cross section and due to the fact that the flux of
neutrinos is slightly higher than that of antineutrinos,
$\nu_\mu/\bar\nu_\mu \sim 1.15$ and $\nu_e/\bar\nu_e \sim 1.25$, for
the energies of interest here~\cite{Barr:2004br, Battistoni:2002ew, Honda:2011nf, Athar:2012it}.

A general feature in both panels is the absence of significant matter
effects for $\cos\theta \gtrsim -0.4$ ($L < 5100$~km), as expected (see Sec.~\ref{subsec:probabilities}).  For neutrinos crossing the Earth deeply, resonant $\tet$-driven matter effects are very important and there are clear differences in the number of events between neutrino-NH and neutrino-IH (antineutrino-NH and antineutrino-IH), and between neutrino-NH (antineutrino-IH) for a density profile given by the PREM and for the case of increasing the density by 10\%.  In particular, notice that there are important differences in the total number of muon-like events in the angular interval $-0.8 < \cos\theta < -0.6$ ($7645~\textrm{km} < L < 10194$~km) for the two cases of the depicted Earth's density, being always more important for NH as it is in this case when the neutrino channel is mostly affected by matter effects.  The larger the density the lower the resonant energy and the shorter the baseline of the first oscillation maximum.  Hence, the crossing of the number of events of these two cases (e.g., the red solid and green dotted lines) for $-0.6 < \cos\theta < -0.5$ is due to the interplay between these two effects.  On the other hand, the matter effect on the propagation of core-crossing neutrinos is smaller than for just-mantle-crossing neutrinos.  This  is due to the resonance energy for the former being outside the energy range we consider here, and hence the effect is due to non-resonant matter effects that modify the propagation in a less significant manner. 

The matter effect is larger in the lower part of the energy range we
consider, as is clear from both panels and was discussed above.  In
general, the resonance energy for just-mantle-crossing neutrinos lies
within the 5-10~GeV energy bin, whereas in the 10-15~GeV energy bin
the effects are non-resonant for all trajectories.  The minimum at
$\cos\theta \sim - 0.5$ for 10-15~GeV (right panel) is explained by
the minimum in the $P(\nu_\nu \rightarrow \nu_\mu)$ probability
(upper-right plot in Fig.~\ref{fig:probability}), which moves to
smaller values of $\cos\theta$ (longer baselines) for higher energies.
Likewise, the minimum at $\cos\theta \sim - 0.3$ for 5-10~GeV is
explained by the first minimum in the $P(\nu_\nu \rightarrow \nu_\mu)$
probability (upper-left and upper-right plots in Fig.~\ref{fig:probability}).  The fact that the second minimum in this
probability does not translate into a minimum in the number of
muon-like events in the energy range 5-10~GeV is because the first
maximum contributes significantly as the full energy range is
integrated. 

Overall, the best sensitivity to the neutrino mass hierarchy
(difference between the red solid and blue dotted lines, and between
the magenta dotted and black dotted lines) and to the matter density
(difference between the red solid and green dotted lines, and between
magenta dotted and brown dotted lines) is achieved in the 5-10~GeV
energy range and for neutrino trajectories such that $-0.8 <\cos\theta
< -0.4$, although some sensitivity remains for other angular and
energy bins.  In this case, the resonant matter effects take place,
whereas for higher energies or for neutrinos traversing the core,
non-resonant effects occur.  As expected, it is clear from both figures that the sensitivity to the mass hierarchy is better than that to changes in the Earth's density.

\subsection{Statistical analysis}

\begin{table}[t]
\begin{center}
\begin{tabular}{||c||c||c||} \hline\hline
\multicolumn{1}{||c||}{{\rule[0mm]{0mm}{6mm}{True Values}}}
& \multicolumn{1}{c||}{\rule[-3mm]{0mm}{6mm}{Marginalization Range}}
& \multicolumn{1}{c||}{\rule[-3mm]{0mm}{6mm}{External $1 \sigma$
    error}} 
\cr
\hline \hline
$\scht = 0.025$ & [0.019, \,0.030] & $\sigma(\stch)=5\%$ \cr 
\hline
$\sat = 0.5$ & [0.38, \, 0.66] & $\sigma(\sta)=2\%$ \cr 
\hline
$\mefft = \pm 2.4 \times 10^{-3} \ {\rm eV}^2$ & 
$[2.2, \, 2.6] \times 10^{-3} \ {\rm eV}^2$ (NH) & $\sigma(|\meff|)= 4\%$ \cr  
& $- [2.6, \, 2.2] \times 10^{-3} \ {\rm eV}^2$ (IH) &  \cr 
\hline
$\mst = 7.62 \times 10^{-5} \ {\rm eV}^2$ & -- & -- \cr 
\hline
$\ssst = 0.32$ & -- & -- \cr
\hline
$\dcpt = 0^{\circ}$ & -- & -- \cr
\hline
$\Delta \rho^{\textrm{true}} = 0$ & [-0.1, \, 0.1] & -- \cr
\hline
$\xi_{\textrm{norm}}^{\textrm{true}} = 0$ & [-1, \, 1] &
$\sigma_{\textrm{norm}} = 20\%$ \cr
\hline \hline
\end{tabular}
\mycaption{\label{tab:benchpar} {\sl \textbf{Benchmark parameters used in the analyses.}  The first column shows the central values of the oscillation parameters used in this work as true values.  The second column shows the ranges of the parameters over which the marginalizations have been performed in the fit ($2\sigma$~CL ranges~\cite{Tortola:2012te}).  Estimated future $1\sigma$~CL errors on the oscillation parameters are given in the last column.  In the last two rows, $\Delta \rho$ is the correction to the normalization of the Earth's matter density with respect to the PREM profile and $\xi_{\textrm{norm}}$ represents the nuisance parameter used to incorporate a fully correlated systematic error ($\sigma_{\textrm{norm}}$) in the normalization of the number of events.}
}
\end{center}
\end{table}

We have generated our simulated data for the true (central) values of
the oscillation parameters given in the first column of Tab.~\ref{tab:benchpar}.  The numbers of simulated muon-like events in each bin for 10~years of data taking in DeepCore and PINGU-0 are
given in Tabs.~\ref{tab:benchDC} and~\ref{tab:benchP0} in Appendix~\ref{app:tables}.  We also show in these tables the results after varying the normalization of the matter density.

The values of the oscillation parameters we use to estimate the
sensitivities lie within the allowed ranges at $1\sigma$~CL
obtained from global fits of the current neutrino data, except for the
value of $\sa$~\cite{Tortola:2012te, Fogli:2012ua, GonzalezGarcia:2012sz}.  Recent global analyses prefer a non-maximal
value of $\tmt$ at about the $2\sigma$~CL which is mainly due to
the recent MINOS $\numu$ disappearance data~\cite{Nichol:2012},
although the new Super-Kamiokande atmospheric neutrino data still
favors maximal 2-3 mixing as the best-fit value in a 2-neutrino
oscillation analysis~\cite{Itow:2012}.  In our analysis, we choose to
use $\sat = 0.5$ as the reference
value.  For the atmospheric neutrino mass splitting, we take an
effective mass-squared difference of $(\meff)^{\textrm{true}} = \pm \, 2.4 \times 10^{-3} \ {\rm eV}^2$ as measured by the accelerator experiments in the $\numu \to \numu$ disappearance channel~\cite{Nichol:2012}.  The $+$ ($-$) sign refers to the case where NH (IH) is the true hierarchy.  This effective mass-squared difference is related to the $\ma$ and $\ms$ mass-squared differences through the expression~\cite{deGouvea:2005hk, Nunokawa:2005nx}
\be
\meff = \ma - \ms (\cos^2\tem - \cos\dcp \, \sin\tet \, \sin2\tem \, \tan\tmt) \, ,
\label{parkedef}
\ee
where $\Delta m^2_{ij}=m^2_i-m^2_j$.  Since the degenerate solutions occur at $\pm \meff$, which is slightly different from $\pm \ma$, it is important to take into account this feature when evaluating the sensitivity to the neutrino mass hierarchy.

In all fits, we marginalize over $\stch$, $\sta$, and $|\meff|$ within their presently allowed $\pm\,2\sigma$ ranges (see the second column of Tab.~\ref{tab:benchpar}).  We impose a prior (Gaussian error at $1\sigma$) of 5\% on $\stch$ considering the expected precision by the Daya Bay experiment by 2016~\cite{dayabay_NF12}.  We also take expected priors of 2\% and 4\% on $\sta$ and $|\meff|$, respectively~\cite{Itow:2001ee}.  We keep $\tem$, $\ms$ and $\dcp$ fixed both in data and in theory and no marginalization has been performed for these oscillation parameters since they would affect our results in a negligible way.  Finally, we also add a fully correlated systematic error in the normalization of the number of events, which could be due to errors on the normalization of the atmospheric neutrino flux, the detector effective mass, the cross section or the efficiency.  We set this error to $\sigma_{\textrm{norm}} = 0.2$.

The treatment of correlated systematic errors is performed by the
Lagrange multiplier method or pull approach~\cite{Brock:2000ud, Pumplin:2000vx, Stump:2001gu, Fogli:2002pt}.  Thus, we consider nuisance parameters that describe the systematic error of the normalization of the number of events ($\xi_{\textrm{norm}}$) and the
errors of $\stch$ ($\xi_{\stch} \equiv \stch - \stcht$), $\sta$ ($\xi_{\sta} \equiv \sta - \stat$) and $|\meff|$ ($\xi_{|\meff|} \equiv |\meff| - |(\meff)^{\textrm{true}}|$).  The variation of these nuisance parameters in the fit is constrained by adding a quadratic penalty to the corresponding $\chi^2$ function without systematic errors.

Let the number of muon-like events detected in the $i$-th angular and
$j$-th energy bin be $N^{\textrm{data}}_{ij}~=~N^{\textrm{th}}_{ij} (\vec{\lambda}^{\textrm{true}})$ and $N^{\textrm{th}}_{ij} (\vec{\lambda})$ the expected number of events in that bin, where 
\be
\vec{\lambda} = \{\tet, \, \tmt,  \, |\meff|, \, h, \, \Delta \rho; \,
\temt, \, \mst, \, \dcpt\} \, ,
\label{eq:lambdarho}
\ee
with $h= \textrm{sign}(\meff)$ and $\Delta \rho$ a global normalization factor for the Earth's density so that $\rho = (1 + \Delta \rho) \, \rho_{\textrm{PREM}}$, with $\rho_{\textrm{PREM}}$ being the PREM density profile.  We estimate the difference between data and theory with the following $\chi^2$ expression~\cite{Brock:2000ud, Pumplin:2000vx, Stump:2001gu, Fogli:2002pt}:
\be
\chi^2 (h, \, \Delta \rho) = 
{\textrm{min}_{(\xi_{\textrm{norm}}, \, \xi_{\stch}, \, \xi_{\sta}, \, \xi_{|\meff|})}} \, \left\{\chi^2_{\textrm{stat}} (\vec{\lambda}, \, \xi_{\textrm{norm}}) \, + \, \left(\frac{\xi_{\textrm{norm}}}{\sigma_{\textrm{norm}}}\right)^2 \, +
\, \chi^2_{\textrm{prior}}\right\} \, ,
\label{eq:chisqrho}
\ee
where min$_{(\xi_{\textrm{norm}}, \, \xi_{\stch}, \, \xi_{\sta}, \, \xi_{|\meff|})}$ indicates the minimum with respect to those parameters.  The $\chi^2_{\textrm{stat}}$ function is given by
\be
\chi^2_{\textrm{stat}} (\vec{\lambda}, \, \xi_{\textrm{norm}}) =
\sum_{(\cos\theta)_i} \, \sum_{(E_{\nu})_j} 
\frac{\left[N_{ij}^{\textrm{data}} - N_{ij}^{\textrm{th}}
    (\vec{\lambda}) (1+\xi_{\textrm{norm}})
    \right]^2}{N_{ij}^{\textrm{data}}} 
\label{eq:chisqstatrho}
\ee
and $\chi^2_{\textrm{prior}}$ is defined as
\be
\chi^2_{\mathrm{prior}} =  \left(\frac{\xi_{\stch}}{\sigma(\stch)}\right)^2 + \left(\frac{\xi_{\sta}}{\sigma(\sta)} \right)^2 + \left(\frac{\xi_{|\meff|}}{\sigma(|\meff|)} \right)^2  \, .
\label{eq:prior}
\ee

In order to estimate the sensitivity to the global normalization of
the Earth's matter density, in the simulated experimental data we set
$\Delta \rho^{\textrm{true}} = 0$, and in the fit we vary $\Delta \rho$ in the range $[-0.1, \, 0.1]$.  Then, we marginalize over the two possibilities for the neutrino mass hierarchy ($h=\pm$), so that we obtain $\chi^2 (\Delta \rho)$.  The sensitivity is calculated as
\be
S(\Delta \rho) = \sqrt{\chi^2 (\Delta \rho) - \chi^2 (0)} \equiv
\sqrt{\Delta \chi^2 (\Delta \rho)} \, .
\label{eq:sensrho}
\ee
We present our results for both NH ($h^{\textrm{true}} = +$) and IH
($h^{\textrm{true}} = -$) as the true mass hierarchy. 

In the case of the determination of the neutrino mass hierarchy, we
define
\be
\chi^2_{\pm} (\Delta \rho) = {\textrm{min}_{(\xi_{\textrm{norm}}, \, \xi_{\stch}, \, \xi_{\sta}, \, \xi_{|\meff|})}} \, \left\{\chi^2_{\textrm{MH}\pm} (\vec{\lambda}_\mp, \, \xi_{\textrm{norm}}) \, + \, \left(\frac{\xi_{\textrm{norm}}}{\sigma_{\textrm{norm}}}\right)^2 \, +
\, \chi^2_{\textrm{prior}}\right\} \, ,
\label{eq:chisqMH}
\ee
for true NH ($\chi^2_+ (\Delta \rho)$) and true IH ($\chi^2_- (\Delta
\rho)$), where 
\be
\chi^2_{\textrm{MH}\pm} (\vec{\lambda}_\mp, \, \xi_{\textrm{norm}}) =
\sum_{(\cos\theta)_i} \, \sum_{(E_{\nu})_j} 
\frac{\left[N_{ij}^{\textrm{th}}(\vec{\lambda}^{\textrm{true}}_\pm) -
    N_{ij}^{\textrm{th}} (\vec{\lambda}_{\mp}) (1+\xi_{\textrm{norm}})
    \right]^2}{N_{ij}^{\textrm{th}}(\vec{\lambda}^{\textrm{true}}_\pm)}
\, ,
\label{chisqstatMH}
\ee
with $\vec{\lambda}_+$ ($\vec{\lambda}_-$) for NH (IH) defined as
\be
\vec{\lambda}_{\pm} = \{\tet, \, \tmt,  \, |\meff|, \, \pm, \, \Delta \rho; \, \temt, \, \mst, \, \dcpt\} \, .
\label{eq:lambdaMH}
\ee
Then we marginalize over $\Delta \rho$ in the range $[-0.1, 0.1]$
to incorporate the impact of possible fluctuations in the global
normalization of the density of all the layers of the PREM profile (but without imposing any prior) to obtain $\chi^2_h$.  The sensitivity in this case is just given by
\be
S_h = \sqrt{\chi^2_h} \, .
\label{eq:sensMH}
\ee

Like for the sensitivity to the Earth's density, we also study both
cases for the neutrino mass hierarchy, when data is generated assuming NH (IH) as the true neutrino mass hierarchy and we compare with the
theoretical expectation when considering IH (NH).

Notice that for both analyses, only the few bins where the $\tet$-driven matter effects are present (see Fig.~\ref{fig:events} and Tabs.~\ref{tab:benchDC} and~\ref{tab:benchP0} in Appendix~\ref{app:tables}) would contribute to the $\chi^2$.  Thus, correlated systematic errors are expected to have a much smaller effect than uncorrelated systematic errors (see
Appendix~\ref{app:note}), as the values of the parameters related to
them get fixed by the bins where the matter effect is negligible.  We
have explicitly checked this for both analyses.

\section{Results}
\label{sec:results}

In this section we show the sensitivities to the Earth's density
distribution and to the neutrino mass hierarchy by using atmospheric
neutrino data for the DeepCore, PINGU-0 and PINGU-I configurations.
We also comment on the impact of systematic errors on the analyses, as
well as the effect of the marginalization over the oscillation
parameters $|\meff|$, $\tet$ and $\tmt$.

\subsection{Sensitivity to the Earth's matter density}
\label{sec:density}

\begin{figure}[t]
\centering
\includegraphics[width=0.5\textwidth]{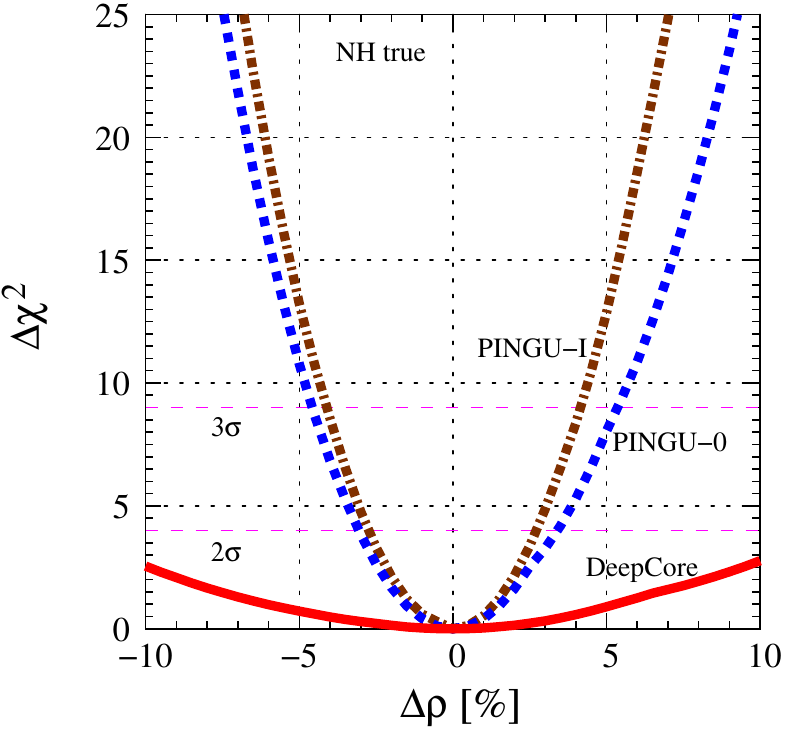}
\includegraphics[width=0.49\textwidth]{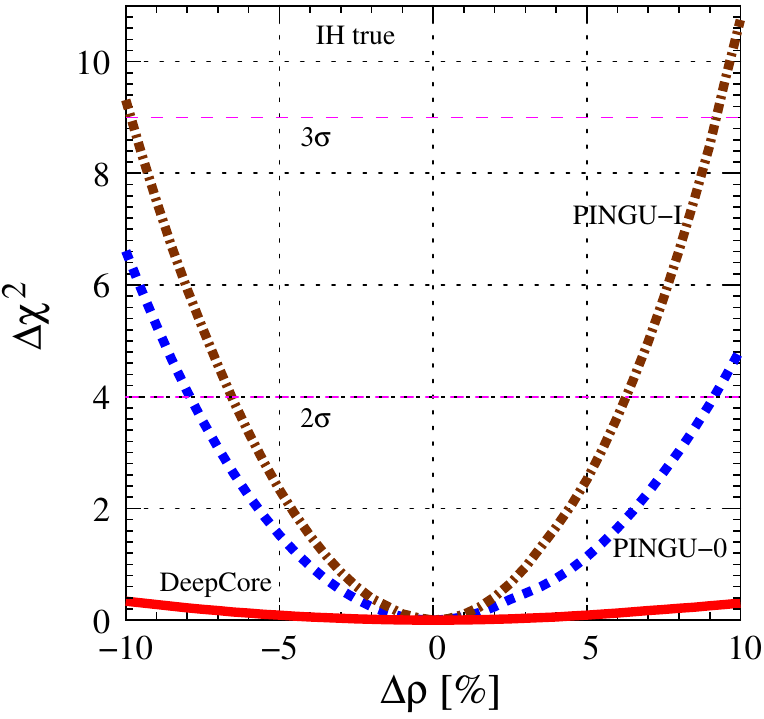}
\mycaption{{\sl \textbf{Sensitivity to fluctuations on the overall normalization of the Earth's density} with respect to the PREM profile, for the DeepCore (red solid lines), PINGU-0 (blue dashed lines) and PINGU-I (brown dash-dotted lines) configurations for the case of NH as the true hierarchy (left panel) and for IH as the true hierarchy (right panel).
}  
\label{fig:alldensity}}
\end{figure}

In Fig.~\ref{fig:alldensity} we show our results for the sensitivity
to the Earth's matter density, for both NH (left panel) and IH (right panel) as the true hierarchy.  We show the sensitivities for the three
detector configurations detailed in Tab.~\ref{tab:detectors} for an
exposure of 10~years.  All the results shown in this section include
the effects of the marginalization, within the current $2 \sigma$~CL
ranges~\cite{Tortola:2012te}, over $\stch$, $\sta$ and $|\meff|$, on which we impose a prior as discussed in the previous section.  We also marginalize over the neutrino mass hierarchy.  Finally, we also include a 20\% fully correlated systematic uncertainty in the total number of muon-like events.  As mentioned above, this systematic error does not have a large impact on the sensitivities due mainly to the presence of the higher $\cos\theta$ (closer to the horizon and thus very little affected by matter effects) angular bins in the analysis, that help to fix the normalization of the number of events in those bins.

For the DeepCore configuration, fluctuations in the normalization of
the Earth's density of $\Delta \rho \simeq \pm 10\%$ can be detected
at the $\sim 1.6\sigma$~CL ($\sim 1\sigma$~CL) in the case where NH (IH) is the true hierarchy.  The prospects are significantly improved for the PINGU-0 and PINGU-I configurations, due to the information contained in the lower energy bins (matter effects are maximal at the resonant energies $\sim 5-7$~GeV, which are unaccessible to the considered DeepCore configuration).  For NH (IH), PINGU-0 would be able to measure matter fluctuations of $\Delta \rho \simeq \pm 3\%$ ($\pm 9\%$) at the $2\sigma$~CL. PINGU-I would improve these numbers
to $\Delta \rho \simeq \pm 2\%$ ($\pm 6\%$) for NH (IH).  

The results are always significantly better for NH.  This can be understood from the fact that the matter effects are present in the neutrino channel, which has more statistics.  For IH the larger
number of events in the neutrino channel (not affected by matter
effects in this case) substantially swamp the $\tet$-driven matter effects in the antineutrino channel. As we have mentioned, these have to be added together, which reduces the sensitivity to density fluctuations, specially in the case of IH.  On the other hand, it is interesting to note that PINGU-0 and PINGU-I are slightly more sensitive to negative fluctuations of the density than to positive ones.  This is because the higher the density, the lower the resonance energy (see lower panel of Fig.~\ref{fig:reso}).  Thus, with an energy threshold of 5~GeV, some of the events affected by the matter effects would lie outside the energy range we consider.  However, for lower densities, more events around the resonance region are included in the data set, which compensates the loss of sensitivity due the marginalizations.  This has a smaller effect in DeepCore since the resonant effects do not take place in the energy bin we consider.  We have also explicitly checked that, for the assumed energy thresholds, these detectors are basically sensitive to fluctuations of the normalization of the Earth's mantle ($L < 10700$~km), as the resonance for cross-crossing neutrinos occurs at lower energies.

Finally, let us comment on the impact of the marginalization over all
the different parameters, which is small.  Indeed, for negative density fluctuations, it is negligible for PINGU-0 and PINGU-I.  For positive fluctuations, the marginalization over $\tmt$ has a small but non-negligible effect for the three detector configurations.  The value of $\sa$ sets the amplitude of the matter effects (see Eqs.~(\ref{eq:P3ee}) - (\ref{eq:P3mutau}) and Eqs.~(\ref{eq:P3eeC}) - (\ref{eq:P3mumuC})) and therefore, in principle, one expects its marginalization to have some impact.  As a consequence of what was just discussed above, the effect would be larger for the case of positive fluctuations for which the sensitivity is slightly worse.  The configuration for which the largest effects of the marginalizations occurs is PINGU-0 (at the level of $\lesssim 10\%$).  DeepCore does not include the resonance region and hence, the effects of the marginalizations are similar for negative and positive fluctuations.  On the other hand, PINGU-I has more energy bins, which helps to distinguish density effects from changes in the mixing parameters.

Notice that the measurement of the matter density profile is much more
difficult than the extraction of the neutrino mass hierarchy, which we
consider next.  This can be explained by the fact that while just
measuring matter effects alone can serve to determine the neutrino
mass spectrum, for the Earth's matter density one also has to be able
to measure perturbations to the matter effects.  These could be small
and difficult to disentangle unless the energy and angular information
of the signal is reasonably good.

\subsection{Sensitivity to the neutrino mass hierarchy}
\label{sec:hierarchy}

\begin{table}[t]
\begin{center}
\begin{tabular}{||c||c||c||c||} \hline\hline
\multicolumn{1}{||c||}{{\rule[0mm]{0mm}{6mm}{Detector configuration}}} 
& \multicolumn{1}{c||}{\rule[-3mm]{0mm}{6mm}{Exposure [years]}} 
&  \multicolumn{1}{c||}{\rule[-3mm]{0mm}{6mm}{CL [$\sigma$] NH(true)}}
&  \multicolumn{1}{c||}{\rule[-3mm]{0mm}{6mm}{CL [$\sigma$] IH(true)}}
\cr
\hline \hline
          & 1  & 2.90  & 2.63 \cr
DeepCore  & 5  & 6.17 & 5.36 \cr
          & 10 & 8.38 & 6.92 \cr
\hline
          & 1  & 4.98 & 4.91 \cr
PINGU-0   & 5  & 10.2 & 8.49 \cr
          & 10 & 14.0 & 11.0 \cr
 \hline
          & 1  & 6.50 & 6.22 \cr
PINGU-I   & 5  & 13.3 & 10.8 \cr
          & 10 & 18.4 & 14.5 \cr
\hline \hline
\end{tabular}
\mycaption{{\sl \textbf{Sensitivity to the neutrino mass hierarchy} for the three detector configurations explored here for different exposure times.  We show the results for NH (third column) and IH (fourth column) as the true hierarchy.}
} 
\label{tab:masshierarchy}
\end{center}
\end{table}

The sensitivity to the neutrino mass hierarchy by exploiting the
number of muon-like events in the three possible neutrino
configurations considered in this work is shown in Tab.~\ref{tab:masshierarchy}, which contains the results of the 
$\chi^2$ analyses for different exposure times (1, 5 and 10~years).
We show the results for the two possible choices of the true neutrino
mass hierarchy, NH and IH.  As in the case of the study of the matter
density fluctuations, for the analysis of the determination of the
neutrino mass hierarchy, we also marginalize over the current $2\sigma$
allowed range of the oscillation parameters $\stch$, $\sta$ and $|\meff|$, imposing the same priors expected from future measurements, and add a 20\% fully correlated systematic error related to the overall normalization of the number of events.  In addition, here we also marginalize over the Earth's matter density distribution, by allowing its normalization to vary freely within $\pm 10\%$.

From the results in Tab.~\ref{tab:masshierarchy} we can see that the
prospects of measuring the neutrino mass hierarchy with these three
detector configurations are very promising. The ongoing DeepCore detector with the signal efficiency and the energy reconstruction capabilities assumed here could provide a $3\sigma$ ($5\sigma$) measurement of the neutrino mass hierarchy after slightly more than 1~year (less than 5~years) if nature has chosen NH and after less than 2~years (5~years) for IH.  As we have already mentioned, the results are worse for the IH scenario because in this case the resonant behavior occurs in the antineutrino channel, which is statistically suppressed due to the smaller antineutrino cross sections and initial fluxes.  On the other hand, with detector configurations such as those considered here for PINGU-0 and PINGU-I, the mass hierarchy could be measured with an astonishing precision, $\sim 5\sigma$ and $>6\sigma$ after 1~year for both NH and IH, respectively.

We have checked that our results agree with previous findings in the
literature~\cite{Mena:2008rh, Akhmedov:2012ah}.  However, note that
those analyses differ on a number of important points from the one
presented here.  In particular, in the recent Ref.~\cite{Akhmedov:2012ah}, a larger effective volume for PINGU was used and no post-trigger efficiency was considered.  Overall this amounts to about a factor of four more statistics in their case.  In that work, a much finer binning, both in $E_\nu$ and $\cos\theta$, was also considered.  This is mainly the reason why they get a sensitivity to the mass hierarchy for 5~years of $45.5\sigma$~CL for NH for the case of no systematic errors and no smearing, whereas for PINGU-0 (PINGU-I), we get $10.2\sigma$~CL ($13.3\sigma$~CL).  On the other hand, whereas we consider large bins for the neutrino energy and zenith angle to approximately take into account the uncertainties in their reconstruction which smear the signal, in Ref.~\cite{Akhmedov:2012ah} these uncertainties are studied more explicitly.  Even with the much larger bins we consider, we would expect our results to slightly worsen when adding this information in a more accurate way.  Another important difference is the way in which systematic errors are added.  Whereas they add uncorrelated errors on the normalization of the number of events, we add fully correlated uncertainties.  As briefly discussed in Appendix~\ref{app:note} and checked numerically, this has important consequences when only a few out of all the considered bins contribute to the $\chi^2$, as is the case here.  In the case of uncorrelated systematic errors in the normalization, the sensitivity gets reduced as the systematic uncertainty increases, whereas for correlated systematic errors the sensitivity is not much affected with respect to the case of no systematic errors, thanks to the use of bins where this error gets fixed.  Hence, their results get worse the larger the systematic error is, whereas the impact of the systematic error on our results is very small.  Finally, we also use different central values for the oscillation parameters, although this has a small effect.

Finally, note that the neutrino mass hierarchy sensitivities computed
here are almost independent of the value of the CP violating phase,
$\dcp$. Therefore the mass hierarchy determination for these three detector configurations would be a clean measurement, free of degeneracies, and could serve as an input for other experiments which
focus on CP violation searches, such as the T2K~\cite{Abe:2011sj} or
NO$\nu$A~\cite{Patterson:2012zs} long-baseline neutrino experiments (see Ref.~\cite{Agarwalla:2012bv} for a recent analysis combining the capabilities of both experiments).

\section{Conclusions}
\label{sec:conclusions}

Atmospheric neutrinos produced by the interactions of cosmic rays in
the Earth's atmosphere cover a huge range of energies and baselines,
offering the opportunity to address many open physics questions.  In
this work we have exploited the low energy region, $E_\nu = 5-15$~GeV,
where matter effects could affect the propagation of these atmospheric neutrinos while they pass deep through the Earth.  If $\tet$ is sufficiently large, the transition probabilities $\nu_{\mu} \rightarrow \nu_{e}$ ($\bar{\nu}_{\mu} \rightarrow \bar{\nu}_{e}$) and $\nu_{e} \rightarrow \nu_{\mu (\tau)}$ ($\bar{\nu}_{e} \rightarrow \bar{\nu}_{\mu (\tau)}$) of atmospheric neutrinos become relevant and resonant matter effects would affect the neutrino (antineutrino) oscillation channels if nature has chosen NH (IH).  Hence, the recent measurement of $\tet \sim 9^\circ$ from several reactor neutrino oscillation experiments~\cite{An:2012eh, Ahn:2012nd, Abe:2012tg}, as well as from the accelerator based T2K experiment~\cite{Abe:2011sj}, brings the opportunity to study these $\tet$-driven matter effects. This could allow us not only to determine the neutrino mass hierarchy, but also to infer some general features of the Earth's density profile.  

Indeed, the idea of exploiting matter effects in atmospheric neutrino oscillations to distinguish the type of neutrino mass hierarchy has been extensively explored in the literature~\cite{Banuls:2001zn, Bernabeu:2001jb, Bernabeu:2002tj, Petcov:1998su, Akhmedov:1998ui, Petcov:1998, Akhmedov:1998xq, Chizhov:1998ug, Chizhov:1999az, Chizhov:1999he, Akhmedov:2005yj, Akhmedov:2006hb, Bernabeu:2001xn, GonzalezGarcia:2002mu, Bernabeu:2003yp, PalomaresRuiz:2003kz, PalomaresRuiz:2004tk, Petcov:2004bd, Indumathi:2004kd, GonzalezGarcia:2004cu, Gandhi:2004bj, Huber:2005ep, Choubey:2005zy, Petcov:2005rv, Indumathi:2006gr, Samanta:2006sj, Gandhi:2007td, Gandhi:2008zs, Samanta:2009qw, Samanta:2010xm, GonzalezGarcia:2011my, Blennow:2012gj, Barger:2012fx, Ghosh:2012, Mena:2008rh, Akhmedov:2012ah}.  On the other hand, three different ways to study the density profile of the Earth by using neutrinos have been considered: neutrino absorption tomography~\cite{Volkova:1974, Nedyalkov:1981, Nedyalkov:1981yy, Nedyalkov:1983, DeRujula:1983ya, Wilson:1983an, Askarian:1985ca, Volkova:1985zc, Tsarev:1985, Borisov:1986sm, Tsarev:1986xg, Kuo:1995, Crawford:1995, Jain:1999kp, Reynoso:2004dt, GonzalezGarcia:2007gg, Borriello:2009ad, Takeuchi:2010, Romero:2011zzb}, neutrino oscillation tomography~\cite{Ermilova:1986ph, Nicolaidis:1987fe, Ermilova:1988pw, Nicolaidis:1990jm, Ohlsson:2001ck, Ohlsson:2001fy, Winter:2005we, Minakata:2006am, Tang:2011wn, Arguelles:2012nw, Ioannisian:2002yj, Akhmedov:2005yt, Lindner:2002wm} and neutrino diffraction~\cite{Fortes:2006}.  These techniques have been discussed for accelerator, extra-terrestrial, atmospheric, solar and supernova neutrinos.  With regards to exploiting the atmospheric neutrino flux, only neutrino absorption tomography had been studied in some detail.  Here we have considered neutrino oscillation tomography with atmospheric neutrinos.

Obtaining a reliable estimate of the density of the Earth is essential to solve a number of important problems in geophysics.  In this work we have presented a completely different technique to determine the Earth's density profile to those used in geophysics, which are mainly based on the analysis of the velocities of seismic waves and on empirical relations between these velocities and density.  Although, in principle, geophysics can obtain more precise results, its inferences are based on numerous assumptions as there are significant trade-offs with other parameters.  Thus, the results from atmospheric
neutrino tomography would represent an independent and complementary
assessment of the Earth's internal structure. 

The goal of this work is the study of the Earth matter effects taking place for atmospheric neutrinos in the GeV range.  We have explored the possibility to do neutrino oscillation tomography to infer information on the Earth's matter density and have studied future sensitivities to determine the neutrino mass hierarchy.  In order to do so, we have considered kilometer-scale ice \v{C}erenkov detectors such as the ongoing DeepCore extension of IceCube~\cite{Collaboration:2011ym, IceCube:2011ah, Ha:2012np}, and PINGU~\cite{Koskinen:2011zz}, which is a further proposed extension to IceCube.  Although these multi-Mton detectors have no charge-identification capabilities that would allow us to distinguish neutrinos from antineutrinos and hence increase the sensitivity to matter effects, the different cross sections for neutrinos and antineutrinos (and also the slightly different fluxes), the large statistics which can be accumulated by these detectors, and the large value measured for the mixing angle $\tet$, make the study of these matter effects feasible and make possible the determination of the neutrino mass hierarchy in relatively short time scales.  

In this work, three different experimental scenarios have been examined (see Tab.~\ref{tab:detectors}).  For the simulations of the DeepCore set-up  we have assumed a single energy bin of $E_\nu = [10, \, 15]$~GeV and an effective mass of $\sim$5~Mton.  For the proposed PINGU experiment, we have considered two possible configurations.  For the PINGU-0 set-up, we have assumed two energy bins: $E_\nu = [5, \, 10]$~GeV and $[10, \, 15]$~GeV and an effective mass of $\sim 7$~Mton.  For the PINGU-I set-up we have assumed four energy bins: $E_\nu = [5.0, \, 7.5]$~GeV, $[7.5, \, 10.0]$~GeV, $[10.0, \, 12.5]$~GeV and $[12.5, \, 15.0]$~GeV and an effective mass of $\sim 7$~Mton.  In addition, in all simulations we have considered a signal efficiency of 50\%.  

In our analyses we have addressed the impact of marginalizations over the neutrino oscillation parameters and we have also included fully correlated systematic uncertainties ($20\%$) on the overall normalization of the total number of events.  Our results are shown in Fig.~\ref{fig:alldensity} and Tab.~\ref{tab:masshierarchy}.  For the current best fit values of the oscillation parameters (see Tab.~\ref{tab:benchpar}), the DeepCore detector, with its relatively high-energy threshold, would only be sensitive to fluctuations on the normalization of the Earth's density of $\Delta \rho \simeq 10\%$ at $\sim1.6\sigma$~CL ($1\sigma$~CL) after 10~years for NH (IH).  On the other hand, the measurement of the neutrino mass hierarchy, being a discrete parameter, is much easier and DeepCore could provide a $>5\sigma$~CL measurement after 5~years of data taking, for both hierarchies.  For the PINGU detector, the lower energy thresholds assumed for the PINGU-0 and PINGU-I set-ups provide better sensitivities to both the Earth's density fluctuations and the neutrino mass hierarchy.  For NH, overall density fluctuations of $\Delta \rho \simeq \pm 3\%$ ($\pm 2\%$) with respect to the PREM profile could be measured at $2\sigma$~CL after 10~years in PINGU-0 (PINGU-I).  In the IH case, the former numbers translate into $\Delta \rho \simeq \pm 9\%$ ($\pm 6\%$).  A measurement of the neutrino mass hierarchy at the $\gtrsim 5\sigma$~CL might be possible within 1~year of data taking, for both hierarchies.  This clearly shows the importance of lowering the energy threshold below 10~GeV, so that detectors are fully sensitive to the resonant $\tet$-driven matter effects.  

In summary, we have studied the future prospects and sensitivities for the ongoing DeepCore and the proposed PINGU detectors to infer overall fluctuations in the Earth's matter density profile and to distinguish the neutrino mass hierarchy. To do this we used muon-like atmospheric neutrino events in the GeV range and exploited the matter effects which affect their propagation through the Earth.  The goal of future multi-Mton detectors must be to lower the energy threshold below 10~GeV so that resonant effects can be accessed which would open the road to address many important physics questions.

\section*{Acknowledgments}

We thank T.~DeYoung and J.~Koskinen for discussions about the performance of DeepCore and PINGU.  SKA and TL acknowledge the support from the Spanish Ministry for Education and Science projects FPA2007-60323 and FPA2011-29678, the Consolider-Ingenio CUP (CSD2008-00037) and CPAN (CSC2007-00042), the Generalitat Valenciana (PROMETEO/2009/116) and the European projects LAGUNA (Project Number 212343).  SKA and OM are supported by the ITN INVISIBLES (Marie Curie Actions, PITN-GA-2011-289442).  OM is also supported by the Consolider Ingenio project CSD2007-00060, by PROMETEO/2009/116 and by the Spanish Grant FPA2011-29678 of the MINECO.  SPR is supported by the Portuguese FCT through CERN/FP/116328/2010 and CFTP-FCT UNIT 777, which are partially funded through POCTI (FEDER) and by the Spanish Grant FPA2011-23596 of the MINECO.

\appendix

\section{Event numbers}
\label{app:tables}

In this appendix we provide some of the event numbers obtained in our analysis, the details of which are given in Sec.~\ref{sec:analysis} in the main text.  Tab.~\ref{tab:benchDC} shows the event numbers relevant for the DeepCore detector set-up and Tab.~\ref{tab:benchP0} those for the PINGU-0 configuration, both for 10~years of data taking.

\begin{table}[h!]
\begin{center}
\begin{tabular}{||c||c||c||c||} \hline\hline
\multicolumn{1}{||c||}{{\rule[0mm]{0mm}{6mm}{
      $\cos\theta$}}} 
& \multicolumn{1}{c||}{\rule[-3mm]{0mm}{6mm}{PREM profile}} 
&  \multicolumn{1}{c||}{\rule[-3mm]{0mm}{6mm}{$\Delta\rho = +10\%$}}
&  \multicolumn{1}{c||}{\rule[-3mm]{0mm}{6mm}{$\Delta\rho = -10\%$}}
\cr
\hline \hline
[-1.0, -0.9] & 2491 (2561) & 2493 (2562) & 2484 (2558)  \cr
\hline
[-0.9, -0.8] & 2087 (2261) & 2124 (2273) & 2067 (2253)  \cr
\hline
[-0.8, -0.7] & 1363 (1589) & 1419 (1606) & 1315 (1572) \cr
\hline
[-0.7, -0.6] & 665 (813) & 669 (814) & 675 (816)  \cr
\hline
[-0.6, -0.5] & 243 (266) & 228 (261) & 261 (273)  \cr
\hline
[-0.5, -0.4] & 279 (225) & 274 (224) & 281 (225)  \cr
\hline
[-0.4, -0.3] & 919 (857) & 925 (860) & 912 (853)  \cr
\hline
[-0.3, -0.2] & 2170 (2133) & 2175 (2136) & 2165 (2130)  \cr
\hline
[-0.2, -0.1] & 3814 (3801) & 3816 (3802) & 3813 (3801)  \cr
\hline                  
[-0.1, 0.0] & 5294 (5292) & 5294 (5292) & 5294 (5292)  \cr
\hline \hline
\end{tabular}
\mycaption{\label{tab:benchDC} {\sl \textbf{
Total number of muon-like contained events after 10~years in the DeepCore detector} in the neutrino energy range 10-15~GeV.  We show the number of events for each angular bin for the PREM density profile (second column) and the cases when the overall normalization is changed by $\pm10\%$ (third and fourth columns).  The numbers without (with) parentheses correspond to NH (IH).  For the oscillation parameters we use the true values given in Tab.~\ref{tab:benchpar} in the main text.}
} 
\end{center}
\end{table}

\begin{table}[h!]
{\small
\begin{center}
\begin{tabular}{|l|c|c|c|c|c|c|}
\hline
\hline
\multicolumn{1}{|c|}{{\rule[0mm]{0mm}{6mm}{{$\cos\theta$}}}}
&\multicolumn{2}{c|}{PREM profile}
&\multicolumn{2}{c|}{$\Delta\rho = +10\%$}
&\multicolumn{2}{c|}{$\Delta\rho = -10\%$} 
\cr
\cline{2-7}
\multicolumn{1}{|c|}{}
&\multicolumn{1}{c|}{5 - 10 GeV}
&\multicolumn{1}{c|}{10 - 15 GeV}
&\multicolumn{1}{c|}{5 - 10 GeV}
&\multicolumn{1}{c|}{10 - 15 GeV}
&\multicolumn{1}{c|}{5 - 10 GeV}
&\multicolumn{1}{c|}{10 - 15 GeV} \cr
\hline
        [-1.0, -0.9]
         & 4031 (4230) & 3649 (3751)
         & 4114 (4274) & 3652 (3753)
         & 4164 (4258) & 3638 (3746) \cr
         \hline
         [-0.9, -0.8]
         & 4511 (4598) & 3057 (3312)
         & 4549 (4588) & 3111 (3329)
         & 4425 (4563) & 3028 (3300) \cr
         \hline
         [-0.8, -0.7]
         & 5066 (4815) & 1996 (2327)
         & 5267 (4856) & 2078 (2353)
         & 4766 (4725) & 1927 (2303) \cr
         \hline
         [-0.7, -0.6]
         & 4832 (5227) & 974 (1191)
         & 5162 (5379) & 980 (1192)
         & 4600 (5095) & 989 (1195) \cr
         \hline
         [-0.6, -0.5]
         & 5787 (6473) & 356 (390)
         & 5794 (6518) & 335 (382)
         & 5869 (6464) & 382 (399) \cr
         \hline
         [-0.5, -0.4]
         & 5837 (6196) & 408 (329)
         & 5689 (6149) & 402 (328)
         & 6001 (6257) & 411 (330) \cr
         \hline
         [-0.4, -0.3]
         & 3260 (3193) & 1346 (1255)
         & 3193 (3161) & 1355 (1259)
         & 3320 (3225) & 1336 (1249) \cr
         \hline
         [-0.3, -0.2]
         & 1804 (1589) & 3179 (3124)
         & 1812 (1588) & 3186 (3128)
         & 1794 (1590) & 3171 (3121) \cr
         \hline
         [-0.2, -0.1]
         & 6319 (6202) & 5587 (5568)
         & 6327 (6203) & 5589 (5569)
         & 6311 (6201) & 5585 (5567) \cr
         \hline
         [-0.1, 0.0]
         & 13513 (13496) & 7754 (7751)
         & 13514 (13496) & 7754 (7751)
         & 13513 (13496) & 7754 (7751) \cr
\hline\hline
\end{tabular}
\mycaption{\label{tab:benchP0} {\sl \textbf{
Total number of muon-like contained events after 10~years in the PINGU-0 configuration}, i.e., with two energy bins, 5-10~GeV and 10-15~GeV.  We show the number of events for each angular bin for the PREM density profile (second and third columns) and the cases when the overall normalization is changed by $\pm10\%$ (fourth to seventh columns).  The numbers without (with) parentheses correspond to NH (IH).  For the oscillation parameters we use the true values given in Tab.~\ref{tab:benchpar} in the main text.}
}
\end{center}
}
\end{table}

\section{Note on uncorrelated and correlated errors}
\label{app:note}

Let us qualitatively explain the effects of uncorrelated and correlated errors on the results. Let us consider a total number of bins equal to $n$,  $N_i^{\textrm{data}}$ as the number of events detected in the $i$-th bin and $N_i^{\textrm{th}} (\vec{\lambda})$ as the number of predicted events for the parameter set $\vec{\lambda}$,  with $\sigma_i^{\textrm{stat}} = \sqrt{N_i^{\textrm{data}}}$ and $\sigma_i^{\textrm{sys}} = \sigma_\xi \, N_i^{\textrm{th}} (\vec{\lambda})$ being the statistical and systematic error in the $i$-th bin, respectively, which satisfy $\sigma_i^{\textrm{stat}} \ll \sigma_i^{\textrm{sys}}$. 

For illustration, let us now consider that only the $i= 1, \, ... \, , m$-th bins contribute to the $\chi^2$ and that their contribution is
approximately the same, i.e., $N_i^{\textrm{data}} - N_i^{\textrm{th}}
(\vec{\lambda}) \simeq \Delta N (\vec{\lambda})$, $\sigma_i^{\textrm{stat}} \simeq \sigma^{\textrm{stat}}$ and $\sigma_i^{\textrm{sys}} \simeq \sigma^{\textrm{sys}}$, for $i \leq m$. 

In the case of uncorrelated systematic errors, the $\chi^2 (\vec{\lambda}, \, \sigma_\xi)$ function is 
\be
\chi^2 (\vec{\lambda}, \, \sigma_\xi) = \sum_{i=1}^n
\frac{\left[N_i^{\textrm{data}} - N_i^{\textrm{th}} (\vec{\lambda})
    \right]^2}{(\sigma_i^{\textrm{stat}})^2 +
  (\sigma_i^{\textrm{sys}})^2} \simeq 
\left(\frac{(\sigma_i^{\textrm{stat}})^2}{(\sigma_i^{\textrm{stat}})^2
  + (\sigma_i^{\textrm{sys}})^2}\right) \, \chi^2 (\vec{\lambda}, \,
0) \, . 
\ee
Thus, regardless of how many bins contribute to the $\chi^2$, the
sensitivity gets reduced by a similar amount with respect to the case
of no systematic errors ($\sigma_\xi = 0$). 

On the other hand, in the case of correlated systematic errors, the
$\chi^2 (\vec{\lambda}, \, \sigma_\xi)$ function is~\cite{Brock:2000ud, Pumplin:2000vx, Stump:2001gu, Fogli:2002pt}
\be
\chi^2 (\vec{\lambda}, \, \sigma_\xi) = 
{\textrm{min}_\xi} \, \left\{\sum_{i=1}^n
\frac{\left[N_i^{\textrm{data}} - N_i^{\textrm{th}} (\vec{\lambda})
    (1+\xi) \right]^2}{\sigma_i^2} + \, 
\left(\frac{\xi}{\sigma_\xi}\right)^2 \, \right\} \, ,
\label{eq:chisqgen}
\ee
where min$_\xi \left\{\tilde\chi^2(\vec{\lambda}, \, \sigma_\xi, \,
\xi)\right\}$ indicates the minimum of $\tilde\chi^2(\vec{\lambda}, \,
\sigma_\xi, \, \xi)$ with respect to $\xi$.  This is a quadratic
polynomial in $\xi$, so the minimum can be calculated analytically.
The result is
\be
\chi^2 (\vec{\lambda}, \, \sigma_\xi) = \chi^2 (\vec{\lambda}, \, 0) -
\frac{\sum_{i=1}^n
  \frac{\sigma_i^{\textrm{sys}}}{(\sigma_i^{\textrm{stat}})^2} \,
  (N_i^{\textrm{data}} - N_i^{\textrm{th}} (\vec{\lambda}))}{1 +
  \sum_{i=1}^n
  \left(\frac{\sigma_i^{\textrm{sys}}}{\sigma_i^{\textrm{stat}}}\right)^2}. 
\label{eq:corr}
\ee
Within the approximations we are considering, Eq.~\ref{eq:corr} can be written as
\ba
\chi^2 (\vec{\lambda}, \, \sigma_\xi) & \simeq &  
\chi^2 (\vec{\lambda}, \, 0) -
\frac{(\sigma^{\textrm{sys}})^2}{(\sigma^{\textrm{stat}})^2 + n \,
  (\sigma^{\textrm{sys}})^2} \,  
\left(\sum_{i=1}^m \frac{N_i^{\textrm{data}} - N_i^{\textrm{th}}
  (\vec{\lambda})}{\sigma^{\textrm{stat}}}\right)^2 \nonumber \\ 
& \simeq &  
\chi^2 (\vec{\lambda}, \, 0) -
\frac{(\sigma^{\textrm{sys}})^2}{(\sigma^{\textrm{stat}})^2 + n \,
  (\sigma^{\textrm{sys}})^2} \, \left[\chi^2 (\vec{\lambda}, \, 0) +
  \sum_{i\neq j}^m \frac{(N_i^{\textrm{data}} - N_i^{\textrm{th}}
    (\vec{\lambda})) \, (N_j^{\textrm{data}} - N_j^{\textrm{th}}
    (\vec{\lambda})) }{(\sigma^{\textrm{stat}})^2}\right] \nonumber \\ 
& \simeq & 
\frac{(\sigma^{\textrm{stat}})^2 + (n-m) \,
  (\sigma^{\textrm{sys}})^2}{(\sigma^{\textrm{stat}})^2 + n \,
  (\sigma^{\textrm{sys}})^2} \, \chi^2 (\vec{\lambda}, \, 0) \simeq \,
\left(1 - \frac{m}{n}\right) \, \chi^2 (\vec{\lambda}, \, 0) . 
\ea
Hence, if most of the bins contribute ($m \lesssim n$), then $\chi^2
(\vec{\lambda}, \, \sigma_\xi) \ll \chi^2 (\vec{\lambda}, \, 0)$ and
the sensitivity gets reduced with respect to the case of no systematic error.  However, if only a few bins contribute ($m \ll n$), then $\chi^2 (\vec{\lambda}, \, \sigma_\xi) \simeq \chi^2 (\vec{\lambda}, \, 0)$, so the sensitivity is only slightly affected by the correlated systematic errors.

\small
\bibliography{telescope-references}
\bibliographystyle{utphys}
\end{document}